\documentclass[preprint,jcp,floats,amsmath,amssymb]{revtex4}

\usepackage{epsfig}

\usepackage{graphicx}
\usepackage{dcolumn}
\usepackage{bm}
\usepackage{chemarr}
\usepackage{multirow}
\usepackage{subfigure}
\usepackage{epstopdf}

\usepackage[T1]{fontenc}
\usepackage{color}
\usepackage[english]{babel}
\usepackage{hyperref}
\usepackage{verbatim} 
\usepackage{bbold}
\usepackage{pdfsync}
\usepackage{paralist}
\usepackage{MnSymbol}
\usepackage[version=3]{mhchem}
\usepackage{chemarr}
\usepackage{amsthm}

\DeclareGraphicsExtensions{.png,.gif,.jpg,.pdf,.bmp}

\begin{document}

\title{The complex chemical Langevin equation}

\author{David Schnoerr $^{1,2}$,  Guido Sanguinetti $^{2}$, Ramon Grima $^{1}$}

\affiliation{$^{1}$ School of Biological Sciences, University of Edinburgh, UK \\ $^{2}$ School of Informatics, University of Edinburgh, UK}

\begin{abstract}
The chemical Langevin equation (CLE) is a popular simulation method to probe the stochastic dynamics of chemical systems. The CLE's main disadvantage is its break down in finite time due to the problem of evaluating square roots of negative quantities whenever the molecule numbers become sufficiently small. We show that this issue is not a numerical integration problem, rather in many systems it is intrinsic to all representations of the CLE. Various methods of correcting the CLE have been proposed which avoid its break down. We show that these methods introduce undesirable artefacts in the CLE's predictions. In particular, for unimolecular systems, these correction methods lead to CLE predictions for the mean concentrations and variance of fluctuations which disagree with those of the chemical master equation. We show that, by extending the domain of the CLE to complex space, break down is eliminated, and the CLE's accuracy for unimolecular systems is restored. Although the molecule numbers are generally complex, we show that the ``complex CLE'' predicts real-valued quantities for the mean concentrations, the moments of intrinsic noise, power spectra and first passage times, hence admitting a physical interpretation. It is also shown to provide a more accurate approximation of the chemical master equation of simple biochemical circuits involving bimolecular reactions than the various corrected forms of the real-valued CLE, the linear-noise approximation and a commonly used two moment-closure approximation.  
\end{abstract}

\maketitle

\section{Introduction}

Stochastic simulation of chemical systems, particularly those of biological interest, has become a common means of studying chemical dynamics (for recent reviews, see for example \cite{RaoArkin2002,Wilkinson2009,Gillespie2013}). A popular Monte Carlo method of performing such simulations is the stochastic simulation algorithm (SSA); this is an exact method of generating sample paths whose (marginal) probability distribution is the solution of the chemical master equation (CME), the accepted description of chemical dynamics in well-mixed and dilute conditions \cite{Gillespie2013}. The disadvantage of the SSA is its computational inefficiency stemming from its simulation of each and every reaction in the system and the considerable ensemble averaging needed to obtain statistically representative results. 

An alternative, often used simulation framework is the chemical Langevin equation (CLE) \cite{Gillespie2000}. This consists of a set of coupled stochastic differential equations describing the time evolution of the molecule numbers of each species. 
It can be shown using Ito calculus that the CLE is equivalent to the chemical Fokker-Planck equation (CFPE) \cite{Gardiner2010} in the sense that the moments of the two methodologies are precisely one and the same. For chemical systems composed of purely unimolecular reactions, the CFPE's predictions for the mean concentrations and the variance of the fluctuations are the same as those of the CME. For chemical systems composed of at least one bimolecular reaction, there is a difference between the predictions of the CFPE and of the CME which vanishes in the limit of large molecule numbers \cite{Kurtz,GrimaThSt2011}. However it has been shown that this difference is typically quite small, even for systems characterised by small molecule numbers \cite{GrimaThSt2011} and hence the CLE / CFPE formalisms present an alternative framework of stochastic simulation to the SSA.  

The CLE formalism is however not without its problems. Two major issues to its use are (i) its unphysical prediction of negative molecule numbers, and (ii) the problem of evaluating square roots of negative quantities, which can happen whenever the molecule numbers become sufficiently negative \cite{Szpruch2009,Higham2011}. We term this second problem the {\it break down} of the CLE. A simple method of circumventing these problems is to enforce positivity of the molecule numbers by rejecting moves of the CLE algorithm which reduce the molecule numbers below zero (reflecting boundary conditions). More sophisticated methods involve modifying the drift and noise terms of the CLE \cite{Wilkie2008,Dana2011}.  All of these methods however can only be justified on computational grounds, rather than following from a microscopic argument \cite{Gillespie2000}: it is therefore unclear how these modifications in the boundary conditions affect the accuracy and validity of the CLE as an approximate method to probe stochastic chemical systems. Clarifying these issues, and proposing a novel, more accurate handling of the CLE break down, is the main purpose of this work.

The rest of this paper is organised as follows. In Section II, we summarise the CLE and CFPE frameworks and show by means of two unimolecular system examples that three methods which circumvent the break down of the CLE lead to CLE predictions for the mean concentrations and the variance which disagree with those of the CME. In particular we show that the phenomenon of break down is inherently due to the fact that the drift-diffusion process described by the CLE / CFPE will with finite probability reach regions in the state space where the diffusion matrix of the CFPE is not positive semi-definite, i.e., breaks a fundamental condition for the well-definition of the CLE / CFPE frameworks. In Section III A we show that by extending the domain of the CLE to complex space, one avoids break down while recovering the exactness of the CLE and the CME formalisms for the first two moments of unimolecular systems. Furthermore the moments of the complex CLE are shown to be real-valued at all times and to hence admit a physical interpretation. We also show that the complex CLE can be used to compute power spectra and first exit times. In Section III B we apply the complex CLE to two examples of biological importance and which feature bimolecular reactions: an enzyme-catalysed reaction and a genetic negative feedback loop. In both cases we show that the predictions of the complex CLE are remarkably similar to those of the CME, while noting significant differences between the CME and corrected forms of the real-valued CLE predictions. We finish in Section IV by a summary and discussion of the merits and limits of the new type of CLE vis-a-vis alternative approaches in the literature.  

\section{Break down of the chemical Langevin equation}

\subsection{The chemical Langevin equation}

Consider a system of chemical species $X_i$ where $i = 1,...,N$ that interact via a set of $R$ reactions
\begin{align} 
\label{eq1}
  \sum_{i=1}^N  s_{ij} X_i \ce{ ->[\quad c_j\quad ]} \sum_{i=1}^N r_{ij} X_i, 
   \quad j = 1, \ldots,  R,
\end{align}
where $c_j$ is the rate constant of reaction $j$; these constants are the same as appearing in the deterministic rate equation formulation of kinetics. We define the $N \times R$ stochiometric matrix $S$ with elements $S_{ij}  =  r_{ij} - s_{ij}$. If the system is well-mixed and sufficiently dilute then the state of the system at any time is fully determined by the state vector \cite{Gillespie2013} $\mathbf{n}=(n_1, \ldots, n_N)$, where $n_i$ is the molecule number of species $X_i$, and the time evolution of the joint probability distribution of the $n_i$ is given by the chemical master equation (CME) \cite{Gillespie2013}:
\begin{align}\label{eq2}
  \partial_t P(\mathbf{n},t) 
  & = 
    \sum_{r=1}^R f_r (\mathbf{n} - \mathbf{S}_r) P(\mathbf{n} - \mathbf{S}_r, t) 
    - \sum_{r=1}^R f_r (\mathbf{n}) P(\mathbf{n}, t).
\end{align}
Here $\mathbf{S}_r$ is a vector whose entries correspond to the $r$th column of the matrix $S$, and $f_r (\mathbf{n})$ are the microscopic propensity functions which describe the rate at which reaction $r$ proceeds. The results derived in this article hold for any system with analytic propensity functions. Thus they hold for chemical systems characterised by propensity functions which are polynomials in the variables $n_i$, such as systems composed of unimolecular, bimolecular and trimolecular reactions. All example systems in this work comprise reactions of order two or lower, for which the function $f_r (\mathbf{n})$ takes the following form: (i) a zeroth-order reaction by which a species is input into a compartment of volume $\Omega$ is described by $f_r(\mathbf{n})=\Omega c_r$; (ii) a unimolecular reaction involving the decay of some species $h$ is described by $f_r(\mathbf{n})=c_r n_h$; (iii) a bimolecular reaction between two molecules of the same species $h$ is described by $f_r(\mathbf{n})= c_r n_h (n_h-1) \Omega^{-1}$; (iii) a bimolecular reaction between two molecules of different species, $h$ and $v$, is described by $f_r(\mathbf{n})=c_r n_h n_v \Omega^{-1}$. 

The CME cannot be exactly solved for many problems of interest and hence the need for approximation methods. Kramers and Moyal developed a Taylor expansion of the CME which upon truncation leads to a partial differential equation approximation of the CME \cite{Kramers1940,Moyal1949,GrimaPRE2012}. Neglecting all terms of order larger than two, one obtains the chemical Fokker-Planck equation (CFPE)
\begin{align}\label{eq3}
   \partial_t P(\mathbf{x},t)
   = &
     \Big[- \sum_{i=1}^N \partial_i A_i(\mathbf{x})
     + \tfrac{1}{2} \sum_{i,j=1}^N \partial_i \partial_j
         B_{ij}(\mathbf{x}) \Big] P(\mathbf{x},t), 
\end{align}
where we denote the continuous variable corresponding to species $X_i$ by $x_i$ and $\partial_i$ denotes the partial derivative with respect to $x_i$, $\partial_i = \partial / \partial x_i$. Note that whereas the state variables are discrete molecule numbers $n_i$ in the CME, they are continuous numbers $x_i$ in the CFPE; it has been shown that the differences between the predictions of the two descriptions tend to zero in the limit of large molecule numbers \cite{Kurtz}. The drift vector $\mathbf{A}$ and diffusion matrix $B$ are given by
\begin{align}\label{eq4}
  A_i(\mathbf{x})
  & = 
    \sum_{r=1}^R S_{ir} f_r(\mathbf{x}),  \\
\label{eq4b}
  B_{ij}(\mathbf{x})
  & = 
    \sum_{r=1}^R S_{ir} S_{jr} f_r(\mathbf{x}),
\end{align}
where $B$ is a positive semi-definite $N \times N$ matrix. A general Fokker-Planck equation (FPE) of the form of Eq. \eqref{eq3} corresponds to a Langevin equation of the type \cite{Gardiner2010}
\begin{align}\label{eq6}
  d \mathbf{x} 
  & = 
    \mathbf{A}(\mathbf{x}) dt + C(\mathbf{x}) \mathbf{dW}, \quad \quad C(\mathbf{x}) C(\mathbf{x})^T = B(\mathbf{x}),
\end{align}
where $\mathbf{dW}$ is a multi-dimensional Wiener process. This is the chemical Langevin equation (CLE). Note that the domain of both the CFPE and of the CLE is always (implicitly) assumed to be that of real numbers since these describe the time evolution of the molecule numbers. There are many choices of $C(\mathbf{x})$ corresponding to different factorisations of the matrix $B(\mathbf{x})$; these lead to as many different representations of the CLE. A commonly used choice, following the seminal paper by Gillespie \cite{Gillespie2000}, is $C_{ir}(\mathbf{x}) = S_{ir} \sqrt{f_r(\mathbf{x})}$ which leads to a CLE of the form
\begin{align}\label{eq8}
  d x_i
  & = 
    \sum_{r=1}^R S_{ir} f_r (\mathbf{x}) dt
    + \sum_{r=1}^R S_{ir} \sqrt{f_r (\mathbf{x})} dW_r.
\end{align}
We shall call this the \emph{standard form} of the CLE throughout the rest of the article. In this form the matrix $C$ has the dimension $N \times R$. For one variable systems, a possible alternative form of the CLE is given by
\begin{align}\label{eq8alt}
  d x_1
  & = 
    \sum_{r=1}^R S_{1r} f_r (x_1) dt
    + \sqrt{\sum_{r=1}^R S_{1r}^2 f_r (x_1)} dW,
\end{align}
wherein there is only one noise source as opposed to $R$ noise sources in the standard form of the CLE. 

For a function $g(\mathbf{x})$, one can derive an ordinary differential equation for the time evolution of its expectation value $\langle g(\mathbf{x}) \rangle$ from the CFPE \eqref{eq3} by multiplying the equation with $g(\mathbf{x})$ and integrating over all $\mathbf{x}$ \cite{Gardiner2010}. Equivalently, one can use Ito's formula to derive an equation for the time evolution of $g(\mathbf{x})$ from the CLE in \eqref{eq6} and averaging subsequently \cite{Gardiner2010}. The equations derived from the CFPE and CLE for the moments are identical. In particular, they depend only on $B(\mathbf{x}) = C(\mathbf{x}) C(\mathbf{x})^T$ and are thus independent of the particular choice for $C(\mathbf{x})$. In this sense, the different choices for $C(\mathbf{x})$ are often claimed to be equivalent in the literature \cite{Melykuti2010}. 

In the next two subsections we show that when simulating the CLE \eqref{eq6}, different choices of $C(\mathbf{x})$ are not necessarily equivalent. The standard form of the CLE breaks down in finite time because the concentrations can be driven negative and then some of the noise terms containing square roots over concentrations cannot be computed. We shall refer to this phenomenon as ``CLE break down'' throughout the rest of the article. We show that this problem can, for some simple chemical systems, be avoided by using a different choice for $C(\mathbf{x})$ but that this is not generally possible, i.e., for many systems the break down of the CLE occurs for all possible choices of $C(\mathbf{x})$. These results taken together imply that the domain of the CLE is generally not that of real numbers. 

\subsection{Unimolecular reaction systems}

\subsubsection*{Example (i): Production and decay of a chemical species}

We start by considering the simplest example of a chemical reaction system 
\begin{align}\label{eq8b}
  \varnothing & \xrightleftharpoons[\quad c_2 \quad ]{c_1} X,
\end{align}
where $c_1$ and $c_2$ are the rate constants characterising the reaction. The CLE Eq. \eqref{eq6} is given by
\begin{align}
d x 
  & = 
    (\Omega c_1- c_2 x)dt+ C(x) dW,
\end{align}
where $C(x) C(x)^T=B=\Omega c_1 + c_2 x$. We consider two forms of the CLE: the standard form with $C_{11}=\sqrt{\Omega c_1}$ and $C_{12}=-\sqrt{c_2 x}$ and a possible alternative form where $C(x)=\sqrt{\Omega c_1+c_2 x}$. 
We rescale time as $\tau = t c_2$ and define $k=\Omega c_1/c_2$. Note that rescaling time also rescales the noise terms since from Ito calculus we have $dW(\tau) = \sqrt{d\tau} = \sqrt{c_2 dt} = \sqrt{c_2} dW(t)$ \cite{Gardiner2010}. The two CLEs are then respectively given by 
\begin{align}
\label{eq8c}
  d x 
  & = 
    (k-x)d \tau + \sqrt{k}\,dW_1 - \sqrt{x}\, dW_2,\\
\label{eq8d}
  d x 
  & = 
    (k-x)d\tau + \sqrt{k+x}\,dW_1.
\end{align}

We first consider the standard CLE given by Eq. \eqref{eq8c}. Assume we start with a positive $x>0$ at $\tau=0$. The noise terms can drive the system towards $x=0$. For $x=0$ the second noise term vanishes and the drift becomes positive. However, due to the first noise term, the variable $x$ becomes negative with a finite probability in a finite time interval and the CLE breaks down.

Next consider the alternative form of the CLE as given by Eq. \eqref{eq8d}. This CLE would break down for $x<-k$. However, since the diffusion term vanishes for $x=-k$ and the drift term becomes $2k>0$, the region $x<-k$ is not accessible and this CLE does not break down (note that since one has to numerically integrate the CLE with a finite time-step, break down may still occur, but this purely numeric effect vanishes in the limit of infinitesimally small time steps). 

We note that other alternative forms of the CLE than the one considered are possible. However, it is easy to verify that all other possible choices of $C$ give rise to CLEs for which the argument in the square roots becomes negative for $x < 0$ (as for the standard form of the CLE) or for $x < -k$ (as for the alternate form given by Eq. \eqref{eq8d}). Hence the two cases considered above provide a complete picture of the breakdown phenomenon (the same applies to the alternative forms of CLEs considered in Appendices A, B and C).

We call the implementations of the CLEs in Eq. \eqref{eq8c} and Eq. \eqref{eq8d}, CLE-R1 and CLE-R2, respectively, and simulate them using the standard Euler-Maruyama algorithm \cite{Higham}. For the CLE-R1, we impose a reflecting boundary at $x=0$ to avoid the break down of the CLE for finite times. The simulation parameters are the time step of the Euler-Maruyama algorithm ($\delta \tau$), the time after which steady-state is assumed to be achieved ($\Delta \tau$) and the number of samples ($N$). Moments are calculated from a single time trajectory by averaging over the fluctuating variables at time points $\Delta \tau$, $2 \Delta \tau$, ..., $N \Delta \tau$; this procedure is repeated ten times leading to ten independent estimates for the moments - the average over the estimates and the standard deviation about these averages are what is plotted in the figures. This simulation protocol is followed throughout the rest of the paper. 

\begin{figure}
 \begin{center}
  \includegraphics[scale=0.5]{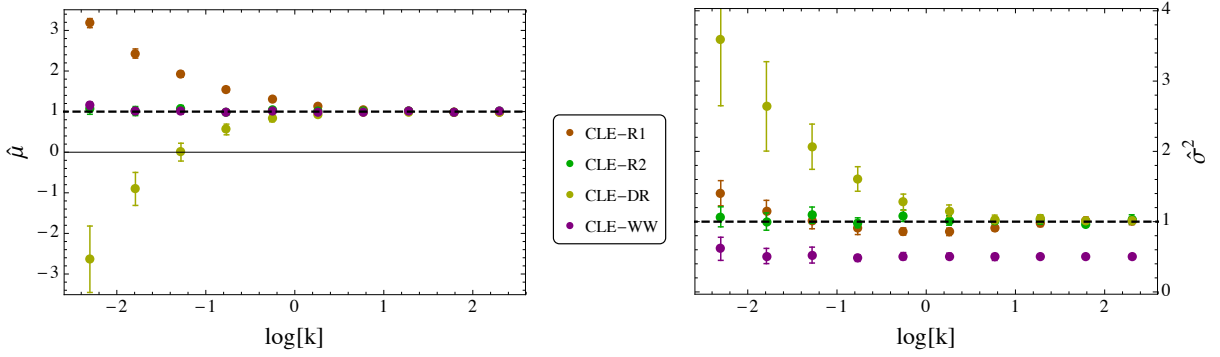} 
  \caption{The normalised mean $\hat{\mu}$ and normalised variance $\hat{\sigma}^2$ as a function of the non-dimensional parameter $k$ for the various CLEs of the simple production-decay reaction system given by scheme \eqref{eq8b}. CLE-R1 is the CLE Eq. \eqref{eq8c} with reflective boundary condition, CLE-R2 is the CLE Eq. \eqref{eq8d}, CLE-WW is the corrected CLE approach in \cite{Wilkie2008} and CLE-DR is the corrected CLE approach in \cite{Dana2011}. The normalisation involves dividing the means and variances obtained from the simulations by the exact analytic results: $\mu=\sigma^2=k =\Omega c_1/c_2$. Only the CLE-R2 agrees with the analytic result (black dashed line) for all $k$.  The simulation parameters are $\delta \tau = 10^{-5}, \Delta \tau = 1, N = 10^3$ (see main text for discussion of these parameters and for the method used to calculate the moments from the CLEs).}
\label{fig1a}
\end{center} 
\end{figure}

Figure \ref{fig1a} shows the results for the mean number of $X$ molecules and the variance of fluctuations about this mean in steady state conditions normalized by the analytic results (mean concentration = variances of fluctuations = $k$ for a birth-death process simulated using the CME or the CFPE) as a function of $k$.  Both methods give the correct result for large values of $k$. This is because a large $k$ value corresponds to a large input-to-decay ratio and thus a large mean value which implies a small probability of the number of molecules becoming negative. With decreasing $k$, the discrepancy between the two CLEs becomes evident: CLE-R1 gives the wrong moments, whereas CLE-R2 agrees with the analytic result. This is clearly due to the fact that CLE-R1 imposes an artificial boundary to avoid the break down of the CLE whereas CLE-R2 naturally does not suffer from any break down. Our results imply that various versions of the CLE are not necessarily equivalent in terms of their boundary behaviour.

Figure \ref{fig1a} shows as well the results of the modified CLE methods of Wilkie and Wong (CLE-WW) \cite{Wilkie2008} and of Dana and Raha (CLE-DR) \cite{Dana2011} applied to the reaction scheme \eqref{eq8b}. The latter method becomes accurate in the macroscopic limit (the limit of large $k$) while the former method (CLE-WW) is accurate only in the mean concentration but gives an incorrect variance of fluctuations for all values of $k$. The CLE-WW does not converge to the correct results in the macroscopic limit because it postulates a global change to the diffusion terms of the CLE (the deletion of some of these terms) to fix the break down problem which is localized to the boundary of zero molecule numbers. On the other hand, the CLE-DR only modifies the drift and diffusion coefficients locally, i.e., when close to the boundary, and hence it necessarily becomes accurate in the macroscopic limit. Because of these reasons, in the rest of this article we shall compare our results only with those of the CLE-DR. 

Hence it is clear that for the simple example considered here, the methods which artificially correct for the break down of the CLE (CLE-R1 with reflection boundary conditions, CLE-DR and CLE-WW) lead to an inequivalence between the CLE's predictions for the first two moments and those of the CME. Equivalence can be restored, in this case, by choosing an alternative CLE representation (CLE-R2) which naturally does not break down at any point in time. We note that the alternative CLE representation is consistent with a drift-diffusion process which can access real values of $x$ larger than $-k$; the probabilistic interpretation of the CFPE is also consistent with such a process since the diffusion scalar $B = \Omega c_1 + c_2 x$ of the CFPE is positive for $x > -k$. Hence one can state that for this example it is possible to find a well-defined CLE representation in real space because the drift-diffusion process describing the chemical reaction lives on the real domain. Similar results as here can be shown for the isomerisation reaction $X_1  \xrightleftharpoons[]{} X_2$ (Appendix A). Next we look at a multispecies unimolecular chemical system, in particular we probe whether one can always find a representation of the CLE in real space which does not suffer break down, i.e, whether the drift-diffusion process associated with general chemical systems inherently lives on the real domain, as conventionally assumed, or not.    

\subsubsection*{Example (ii): Production followed by isomerisation}

We consider the following system of unimolecular reactions involving two distinct species
\begin{align}\label{eq9}
  \varnothing & \xrightleftharpoons[\quad c_4 \quad ]{c_1} X_1 \xrightleftharpoons[\quad c_3 \quad ]{c_2} X_2.
\end{align}
We rescale time as $\tau= c_4 t$ and define $k_1= \Omega c_1/c_4, k_2=c_2/c_4, k_3= c_3/c_4$.
The standard form of the CLE for the reaction system \eqref{eq9} (denoted as CLE-R1) reads
\begin{align}\label{app_bd_1}
  dx_1
  & =
     (k_1 -  k_2 x_1 +  k_3 x_2 - x_1) d\tau 
     + \sqrt{k_1} dW_1 - \sqrt{k_2 x_1} dW_2 + \sqrt{k_3 x_2} dW_3 -  \sqrt{x_1} dW_4, \\
  dx_2
  & =
    ( k_2 x_1 - k_3 x_2) d\tau
    +\sqrt{k_2 x_1} dW_2 - \sqrt{k_3 x_2} dW_3.
\end{align}
When one of the variables becomes zero, some noise terms become zero and some remain finite and thus the noise can drive the system to negative values of the variables which then leads to break down.

A possible alternative form is given by the Langevin equation (denoted as CLE-R2)
\begin{align}
  dx_1
  & =
     (- x_1 +  k_1 -  k_2 x_1 +  k_3 x_2) d\tau + \sqrt{y_1} dW_1 + \sqrt{y_2} dW_2, \\
  dx_2
  & =
    ( k_2 x_1 - k_3 x_2) d\tau
    - \sqrt{y_2}~dW_2,
\end{align}
where we have defined
\begin{align}
  y_1 
  & = 
    x_1+k_1, \\
  y_2
  & = 
    k_2 x_1 + k_3 x_2.
\end{align}
The CLE-R2 breaks down if $y_1$ or $y_2$ become negative. To probe whether this can occur, we transform the CLE-R2 to the new variables $y_1$ and $y_2$. For this purpose, we express $x_1$ and $x_2$ in terms of $y_1$ and $y_2$ as
\begin{align}
  x_1 
  & = 
    y_1 - k_1, \\
  x_2
  & = 
    \frac{1}{k_3} (y_2 + k_2(k_1-y_1)).
\end{align}
Using Ito's formula, it can be shown that the CLE-R2 in the new variables \cite{Gardiner2010} reads 
\begin{align}
  dy_1
   & =
    (2 k_1 - y_1 (2k_2+1) +  2 k_1 k_2 + y_2) d\tau + \sqrt{y_1} dW_1 + \sqrt{y_2} dW_2, \\
  d y_2
   & = 
        (2 k_1 k_2 (1 + k_2 - k_3) + k_2 y_1 (2 k_3 - 2k_2 - 1) + y_2 (k_2 - k_3)) d\tau \notag \\
    & \quad
      + k_2 \sqrt{y_1} dW_1 + (k_2-k_3) \sqrt{y_2} dW_2.
\end{align}
Consider the case $y_2=0, y_1>0$. The CLE-R2 reads
\begin{align}
  d y_2
  & = 
        (2 k_1 k_2 (1 + k_2 - k_3) + k_2 y_1 (2 k_3 - 2k_2 - 1)) d\tau  + k_2 \sqrt{y_1} dW_1.
\end{align}
Clearly the diffusion term can drive the system to negative values of $y_2$ and hence to break down. Interestingly, break down can also occur because the drift becomes negative for positive $y_1$. For example for $2 k_3 - 2k_2 - 1 \neq 0$ and $(1 + k_2 - k_3) / (2 k_3 - 2k_2 - 1) < 0$, the drift becomes negative for $y_1 < 2 k_1 (k_3 - k_2 - 1) / (2 k_3 - 2k_2 - 1)$, which is possible under the constraint $y_1>0$. Similarly it is easy to show that for the case $y_1=0, y_2>0$ the diffusion term can drive the system to break down (break down due to the drift term is here not possible because the drift is always positive). 

To gain insight into the underlying reason for break down, we next consider the diffusion matrix of the CFPE. Using Eq. (\ref{eq4b}) we find the diffusion matrix is given by
\begin{align}
\label{dmatrixCLER}
  B
  & = 
    \left(
	\begin{array}{cc}
	  k_1 + x_1 +  k_2 x_1 +  k_3 x_2 & -  k_2 x_1-  k_3 x_2 \\
 	-  k_2 x_1-  k_3 x_2 &   k_2 x_1+  k_3 x_2 \\
	\end{array}
    \right) 
  = 
    \left(
	\begin{array}{cc}
	  y_1 + y_2 &  - y_2 \\
 	  -y_2  &  y_2  \\
	\end{array}
    \right) .
\end{align}
Its eigenvalues and corresponding eigenvectors are given by
\begin{align}
  \lambda_1 
 & = 
   \frac{1}{2} \left( y_1 + 2 y_2 - \sqrt{y_1^2 + 4 y_2^2} \right), \\
  \lambda_2 
 & = 
   \frac{1}{2} \left( y_1 + 2 y_2 + \sqrt{y_1^2 + 4 y_2^2} \right), \\
  v_1 
  & = 
    \left( - \frac{-y_1 + \sqrt{y_1^2+4 y_2^2}}{2 y_2}, - 1 \right)^T, \\
  v_2
  & = 
    \left( - \frac{y_1 + \sqrt{y_1^2+4 y_2^2}}{2 y_2},  1 \right)^T.
\end{align}
Inspection of these equations shows that the eigenvalue $\lambda_1$ becomes negative if $y_1$ or $y_2$ become negative, i.e., the positive semi-definite form of the diffusion matrix, which is a necessary requirement of any Fokker-Planck equation, cannot be maintained. Hence it follows that break down of CLE-R2 is due to the fact that the drift-diffusion process has a finite probability of accessing a region of space ($y_1$ or $y_2$ negative) where the diffusion matrix of the CFPE is not positive semi-definite, i.e., there is no probabilistic interpretation of a drift-diffusion process in the real domain which describes the reaction system (\ref{eq9}). Therefore, \emph{the break down of CLE-R2 is not due to the particular choice of $C$ underlying it and thus the same problem is manifest for all possible choices of $C$, for all possible Langevin equation representations of the CFPE}. 

\begin{figure} [t]
  \includegraphics[scale=0.4]{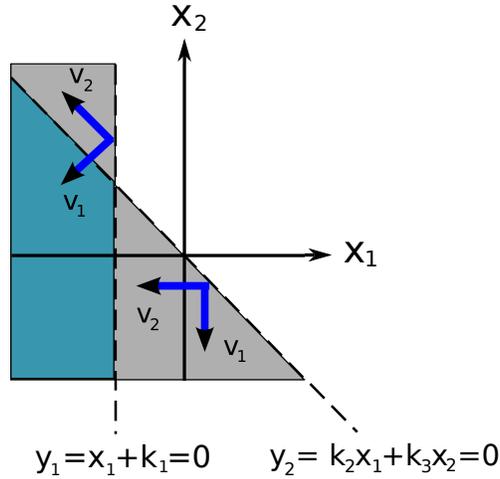} 
  \caption{Graphical representation of the state space of the two-dimensional reaction system in \eqref{eq9}. The dashed lines indicate the boundaries where either $y_1$ or $y_2$ become zero. The grey area corresponds to the part of the state space where the eigenvalues $\lambda_1$ and $\lambda_2$ of the CFPE diffusion matrix are negative and positive, respectively. The blue shaded area represents the region of space where both eigenvalues are negative. Thus the diffusion matrix is not positive semi-definite in the light and blue shaded areas. The blue arrows represent the eigenvectors for the case $y_1=0, y_2>0$ and the case $y_2=0, y_1>0$. Since the eigenvector of the non-vanishing eigenvalue is not parallel to the boundary, there is a non-vanishing noise component orthogonal to the boundary that can drive the system to break down.}
  \label{fig6}
\end{figure}

This intrinsic break down of the CFPE can also be intuited as follows. In Figure \ref{fig6} we show the eigenvectors corresponding to $y_1=0, y_2>0$ and $y_2=0, y_1>0$. The eigenvector corresponding to $\lambda_2$ in the case $y_1=0, y_2>0$ is $v_2 = (-1, 1)^T$. This is clearly not parallel to the boundary $y_1 = x_1+k_1=0$ which is parallel to $(0,1)^T$. There is thus always a non-vanishing noise component orthogonal to the boundary which implies that the noise can drive the system across the boundary thus leading to break down of the CFPE. A similar conclusion follows for the case $y_2=0, y_1>0$. 

We note that the connection between the form of the diffusion matrix and the breakdown properties of the CLE is not specific to this example. It can be generally proved for all chemical systems that if the diffusion matrix $B$ is not positive semi-definite then the matrix $C$ cannot be real, i.e.,~the CLE necessarily breaks down due to square roots of negative arguments. A proof of this result can be found in Appendix E.

\begin{figure}[t]
  \includegraphics[scale=0.5]{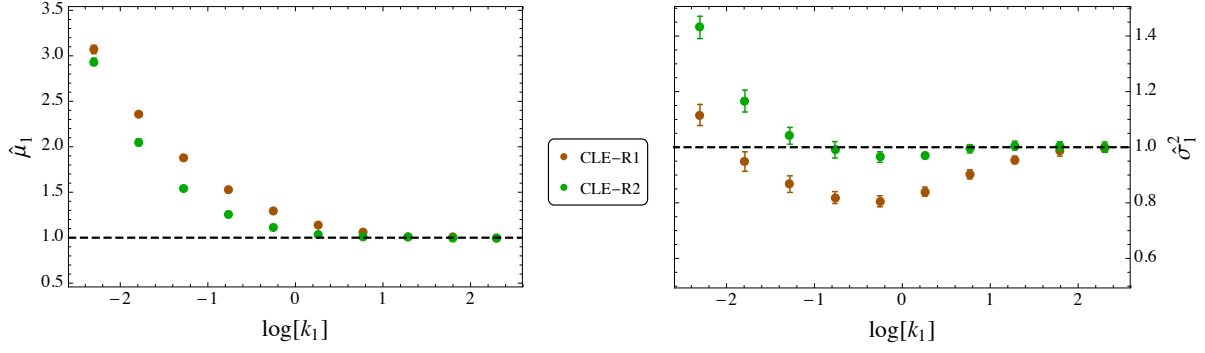}    
  \caption{Normalised mean and variance of the implementations CLE-R1 and CLE-R2 for species $X_1$ of the unimolecular reaction system \eqref{eq9} as a function of $k_1= \Omega c_1/c_4$. All other parameters are set to unity. Reflection boundary conditions at zero molecule numbers and at $y_1=y_2 = 0$ are respectively imposed on CLE-R1 and CLE-R2 to avoid their break down. The values are normalized by the exact analytic expression for the moments obtained from the CME: $\mu_1=\sigma_1^2=k_1$. A large $k_1$ thus corresponds to a large mean value. The dashed line represents the exact value. Since the system is linear, the CLE should reproduce the exact result. The large deviations for small values of $k_1$ thus clearly indicate that the imposed reflection boundaries distort the moments. Simulation details for both implementations are $\delta \tau = 10^{-4}, \Delta \tau = 1, N = 10^4$.}
\label{fig1}
\end{figure}

Correcting the break down by imposing artificial reflective boundaries introduces significant errors. The results of such simulations - the mean and variance of species $X_1$ for CLE-R1 and CLE-R2 - are shown in Figure \ref{fig1}. The results are normalised with the exact analytic results obtained by solving the CME for the reaction system (\ref{eq9}) (this leads to mean = variance = $k_1$). Both CLEs show significant deviations from the exact result for small values of $k_1$, i.e, for small values of the average number of molecules of $X_1$. As for the previous example of production and decay of a chemical species, it is found that these significant deviations from the exact CME result cannot be eliminated using CLEs with modified propensities, i.e., using the methods of Wilkie and Wong \cite{Wilkie2008} and of Dana and Raha \cite{Dana2011}. 

\subsection{Bimolecular reaction systems}

Earlier we saw that for one variable unimolecular systems there is a representation of the CLE which avoids the break down of the standard form of the CLE and which recovers the equivalence of the CLE and CME results for the mean concentrations and variance of fluctuations of unimolecular systems. Contrastingly a break down analysis for one variable systems involving a bimolecular reaction leads to different conclusions: all forms of the CLE can lead to break down in finite time, depending on the initial value of the number of molecules. A detailed analysis of this phenomenon for the two systems of reactions $\varnothing \xrightarrow{} X_1, \quad X_1 + X_1 \xrightarrow{} \varnothing$ and $X_1 + X_1 \xrightleftharpoons[]{} X_2$ can be found in  Appendices B and C, respectively. 

The same conclusion, i.e., the impossibility of fixing break down using an alternative CLE representation, holds also for a wide class of multivariate bimolecular systems. For the latter, break down is independent of the initial conditions, in contrast to what was found for univariate bimolecular systems. The intrinsic reason for the break down is found to be as for unimolecular systems - namely that the diffusion matrix of the associated CFPE loses its positive semi-definite form for points in real number space which can be reached by the drift-diffusion process described by the CLE / CFPE. A detailed break down analysis of the three variable CLE describing a reaction which is catalysed by two enzymes can be found in Appendix D. With the intuitive eigenvector picture in mind (as illustrated in Figure 3), we expect most multi-dimensional systems to break down, since there is no reason why the eigenvectors of the diffusion matrix should in general be parallel to the boundary separating the regions in state space where the diffusion matrix is positive semi-definite and where it is not. 

\section{The complex chemical Langevin equation}

In the previous section we have shown that the commonly employed CLE generally suffers from a break down at finite times due to the occurrence of negative arguments in square roots. This problem can be alleviated by the reflection boundary method or by a variety of other propensity modification methods. However, as we have shown, these procedures introduce inaccuracies in the CLE predictions. Foremost amongst such is the inequivalence between the modified CLE predictions and those of the CME for the mean concentrations and variance of fluctuations of unimolecular systems. 

The state space of the CLE is frequently taken to be the real domain since molecule numbers are real, and this has been an assumption in our derivations in the previous section as well. However as we show in this section, the break down can be avoided by working directly with a complex extension of the CLE; we will show here that this restores the equivalence of the CLE and CME predictions for unimolecular reactions (up to two moments) and gives strikingly accurate results for bimolecular systems. Although the molecule numbers are generally complex, we show that generally the ``complex CLE'' predicts real-valued mean concentrations and moments of intrinsic noise, hence admitting a physical interpretation. 

For clarity, we develop this approach first on the two-species unimolecular system considered earlier, then we extend the latter to the general case and finally present some applications of the complex CLE to two problems of biochemical interest and which involve bimolecular reactions. 

\subsection{An illustrative example}

We consider again the two species system described by scheme \eqref{eq9}. The CLE for this system breaks down independently of the representation, if the state space is real. We now lift this restriction and let the state space be complex, i.e., the CLE now reads
\begin{align}\label{ext_example_eq1}
  dz_1
  & =
     (k_1 -  k_2 z_1 +  k_3 z_2 - z_1) d\tau 
     + \sqrt{k_1} dW_1 - \sqrt{k_2 z_1} dW_2 + \sqrt{k_3 z_2} dW_3 -  \sqrt{z_1} dW_4, \\
  dz_2
  & =
    ( k_2 z_1 - k_3 z_2) d\tau
    +\sqrt{k_2 z_1} dW_2 - \sqrt{k_3 z_2} dW_3,
\end{align}
where $z_1$ and $z_2$ are complex variables. We shall refer to these equations as the CLE-C and to the conventional CLE in real space as the CLE-R. Writing $z_1 = x_1 + i y_1, z_2 = x_2 + i y_2$ where $x_1, x_2, y_1, x_2 \in \mathbb{R}$ and defining the vectors $\mathbf{x}=(x_1,x_2)^T, \mathbf{y}=(y_1,y_2)^T, \mathbf{w} = (x_1, x_2, y_1, y_2)^T$, we can then write stochastic differential equations for the real and imaginary parts as follows
\begin{align}\label{ext_example_eq4}
  d \mathbf{w}
  & = 
    \mathcal{A} dt + \mathcal{C} d \bf{ \mathcal{W}}, 
\end{align}
where we defined 
\begin{align}\label{ext_example_eq5}
  \mathcal{A}
  & =
    \begin{pmatrix}
      A_1^x \\
      A_2^x \\
      A_1^y \\
      A_2^y \\
    \end{pmatrix} 
  =
    \begin{pmatrix}
      k_1 -  k_2 x_1 +  k_3 x_2 - x_1 \\
      k_2 x_1 - k_3 x_2 \\
      -  k_2 y_1 +  k_3 y_2 - y_1 \\
      k_2 y_1 - k_3 y_2
    \end{pmatrix}, \\
  \mathcal{C}
  & = 
    \begin{pmatrix}
      \sqrt{k_1}  &  - \sqrt{k_2} ~\text{Re}(\sqrt{z_1})  &  \sqrt{k_3} ~\text{Re}(\sqrt{z_2})  &   -\text{Re}(\sqrt{z_1}) \\
      0  &  \sqrt{k_2} ~\text{Re}(\sqrt{z_1})  &  - \sqrt{k_3} ~\text{Re}(\sqrt{z_2})  &  0 \\
      0  &  - \sqrt{k_2} ~\text{Im}(\sqrt{z_1})  &  \sqrt{k_3} ~\text{Im}(\sqrt{z_2})  &  -  \text{Im}(\sqrt{z_1}) \\
      0  &  \sqrt{k_2} ~\text{Im}(\sqrt{z_1})  &  - \sqrt{k_3} ~\text{Im}(\sqrt{z_2})  &  0 \\
    \end{pmatrix}, \\
\label{noiseccle}
  d \bf{ \mathcal{W}}
  & = 
    (dW_1, dW_2, dW_3, dW_4)^T.
\end{align}
We use the principal value for the complex square root
\begin{align}\label{ext_example_eq3}
  \sqrt{z_j} 
  & = 
    \sqrt{\frac{\sqrt{x_j^2+y_j^2}+x_j}{2}} + i~ \text{sign}(y_j) \sqrt{\frac{\sqrt{x_j^2+y_j^2}-x_j}{2}}, \quad j =1,2.
\end{align}
The CLE in \eqref{ext_example_eq4} is thus equivalent to the FPE
\begin{align}
   \partial_t P(\mathbf{w},t)
   = &
     \Big[- \sum_i \partial_i \mathcal{A}_i(\mathbf{w})
     + \tfrac{1}{2} \sum_{i,j} \partial_i \partial_j
         \mathcal{B}_{ij}(\mathbf{w}) \Big] P(\mathbf{w},t).
\end{align}
The diffusion matrix of this FPE is given by
\begin{align}\label{ext_example_eq7a}
  \mathcal{B}(\mathbf{w}) 
  & =
    \mathcal{C}(\mathbf{w}) \mathcal{C}(\mathbf{w})^T 
  =
        \begin{pmatrix}
          \mathcal{B}^{xx}  &  \mathcal{B}^{xy}  \\
          \mathcal{B}^{yx}  &  \mathcal{B}^{yy}
    \end{pmatrix}\\
  & = \frac{1}{2}
  \label{ext_example_eq7b}
    \begin{pmatrix}
       2 k_1 + (1+k_2)  p_1 + k_3  p_2  &
           - k_2  p_1 - k_3  p_2  &
           (1+k_2) y_1 + k_3 y_2  &
           -k_2 y_1 - k_3 y_2  \\
       -k_2  p_1 - k_3 p_2  &
           k_2 p_1 + k_3 p_2  &
           -k_2 y_1 - k_3 y_2  &
           k_2 y_1 + k_3 y_2  \\
       (1+k_2) y_1 + k_3 y_2  &
           -k_2 y_1 - k_3 y_2  &
           (1+k_2) m_1 + k_3 m_2  &
           -k_2 m_1 - k_3 m_2  \\
       -k_2 y_1 - k_3 y_2  &
           k_2 y_1 + k_3 y_2  &
           -k_2 m_1 - k_3 m_2  &
           k_2 m_1 + k_3 m_2
    \end{pmatrix},
\end{align}
where we used the definitions $p_{1/2} = \sqrt{x_{1/2} ^2+y_{1/2} ^2}+x_{1/2}$ and $m_{1/2} = \sqrt{x_{1/2} ^2+y_{1/2} ^2}-x_{1/2}$. 

We find that all entries of $\mathcal{B}$ are analytic functions of the $w_i$'s. Moreover, since $\mathcal{B}= \mathcal{C} \mathcal{C}^T$, and $\mathcal{C}$ has real entries then it follows that $\mathcal{B}$ is always positive semi-definite (see Appendix E). In contrast note that the diffusion matrix for the FPE in the real domain, {\it{did not maintain positive semi-definiteness}} for all values of the molecule numbers (see Eq. (\ref{dmatrixCLER}) and the discussion thereafter). 

The next and final question is whether the moments of the complex variables $z_1$ and $z_2$ are real. This is an important question since if this is not the case then the CLE-C does not admit a physical interpretation of the chemical processes it is supposed to describe. To show that this is the case, we first prove invariance of the drift and diffusion operators under a certain operation.

Consider the drift term under the joint reflection of the imaginary parts on the real axes: $\mathbf{y} \to - \mathbf{y}$. Furthermore define 
\begin{align}
\label{ext_example_eq5b}
 \mathcal{A}^{x/y}
 & =
  \begin{pmatrix}
     A_1^{x/y} \\
     A_2^{x/y} \\
   \end{pmatrix}. 
\end{align}
Since $\mathcal{A}^x$ and $\mathcal{A}^y$ are linear in $x$ and $y$, respectively, and independent of the respective other variables, we find $\mathcal{A}^x (\mathbf{x},-\mathbf{y}) = \mathcal{A}^x(\mathbf{x},\mathbf{y})$ and $\mathcal{A}^y (\mathbf{x},-\mathbf{y}) = - \mathcal{A}^y(\mathbf{x},\mathbf{y})$. These combined with $\partial_{-y_i} = - \partial_{y_i}$ imply that the drift operator is invariant under $\mathbf{y} \to -\mathbf{y}$:
\begin{align}
  \partial_{x_i} \mathcal{A}^x (\mathbf{x},-\mathbf{y}) 
  & = 
    \partial_{x_i} \mathcal{A}^x (\mathbf{x},\mathbf{y}) , \\
  \partial_{-y_i} \mathcal{A}^y (\mathbf{x},-\mathbf{y})  
  & = 
    \partial_{y_i} \mathcal{A}^y (\mathbf{x},\mathbf{y}). 
\end{align}
Similarly one can show that the diffusion operator is also invariant under the same operation, as follows. From the definitions of $p_1, p_2, m_1$ and $m_2$ we find that the latter are invariant under the operation $\mathbf{y} \to -\mathbf{y}$. From Eqs. \eqref{ext_example_eq7a} and \eqref{ext_example_eq7b} we find that $\mathcal{B}^{xx}, \mathcal{B}^{xy}$ and $\mathcal{B}^{yy}$ are linear in $(p_1,p_2), (y_1,y_2)$ and $(m_1, m_2)$, respectively ($\mathcal{B}^{yx}$ is equal to $\mathcal{B}^{xy}$). We thus have $\mathcal{B}^{xx}(\mathbf{x},-\mathbf{y}) = \mathcal{B}^{xx}(\mathbf{x},\mathbf{y})$, $\mathcal{B}^{xy}(\mathbf{x},-\mathbf{y}) = -\mathcal{B}^{xy}(\mathbf{x},\mathbf{y})$ and $\mathcal{B}^{yy}(\mathbf{x},-\mathbf{y}) = \mathcal{B}^{yy}(\mathbf{x},\mathbf{y})$ and therefore invariance of the diffusion operator follows
\begin{align}
  \partial_{x_i} \partial_{x_j} \mathcal{B}^{xx}_{ij}(\mathbf{x},-\mathbf{y}) 
  & = 
    \partial_{x_i} \partial_{x_j} \mathcal{B}^{xx}_{ij}(\mathbf{x},\mathbf{y}), \\
  \partial_{x_i} \partial_{-y_j} \mathcal{B}^{xy}_{ij}(\mathbf{x},-\mathbf{y})  
  & = 
    \partial_{x_i} \partial_{y_j} \mathcal{B}^{xy}_{ij}(\mathbf{x},\mathbf{y}) , \\
  \partial_{-y_i} \partial_{-y_j} \mathcal{B}^{yy}_{ij}(\mathbf{x},-\mathbf{y}) 
  & = 
    \partial_{y_i} \partial_{y_j} \mathcal{B}^{yy}_{ij}(\mathbf{x},\mathbf{y}).
\end{align}
Since both the drift and diffusion operators are invariant under the reflection $\mathbf{y} \to -\mathbf{y}$, it follows that the whole FPE is invariant as well. Now the initial condition is always such that the imaginary part $\mathbf{y}$ is zero which implies that the probability distribution is initially symmetric in $\mathbf{y}$; since the FPE is invariant under the reflection $\mathbf{y} \to -\mathbf{y}$, one is led to the conclusion that the probability distribution solution of the FPE for all times has the property: $P(\mathbf{x}, \mathbf{y}, t) = P(\mathbf{x},-\mathbf{y},t)$. This in turn will allow us to show that the moments of the complex variables $z_i$ are real, as follows. 

Consider now a general moment $\langle z_1^{m_1} z_2^{m_2} \rangle, m_1, m_2 \in \mathbb{N}$,  of the complex variables $z_i=x_i + i y_i$
\begin{align}\label{mtext_con_eq3}
  \langle z_1^{m_1} z_2^{m_2} \rangle
  & =
    \int dz_1 dz_2~
    z_1^{m_1} z_2^{m_2} P(\mathbf{z},t) \notag \\
  & =
    \int dx_1 dx_2 dy_1 dy_2~
    (x_1+ i y_1)^{m_1} (x_2 + i y_2)^{m_2} P(\mathbf{x},\mathbf{y},t).
\end{align}
Each summand of the imaginary part of the product $(x_1+ i y_1)^{m_1} (x_2 + i y_2)^{m_2}$ is proportional to $x_1^{m_1-k_1} x_2^{m_2-k_2} y_1^{k_1} y_2^{k_2}$, with $k_i \in \mathbb{N}, k_i \leq m_i$ for $i=1, 2$, and $\sum_{i=1}^2 k_i$ is odd, i.e.~the exponents of the $y_i$ sum to an odd integer. The term $x_1^{m_1-k_1} x_2^{m_2-k_2} y_1^{k_1} y_2^{k_2}$ is thus an odd function in $\mathbf{y}$; since the probability distribution is symmetric in $y$ it then follows that the integral over the imaginary part in Eq.~\eqref{mtext_con_eq3} is equal to zero. Hence the \emph{moments of the complex variables $z_i$ are real at all times}.

We simulated the complex CLE (CLE-C) for scheme \eqref{eq9} to (i) verify that the moments of its complex variables are real and (ii) test the accuracy of its predictions versus those obtained using the real-valued CLE (standard and alternative forms as considered in Section II B) with reflective boundary conditions. The results are shown in Figures 4 and 5, respectively. The simulations of the CLE-C consist in the simulation of four coupled stochastic differential equations for the real and imaginary parts of the complex concentrations of species $X_1$ and $X_2$; these are given by Eqs. (\ref{ext_example_eq4})-(\ref{noiseccle}). We find that for all five moments, the imaginary part scales as $N^{-1/2}$, where $N$ is the number of simulated samples (see Figure 4); this law strongly suggests that the non-zero value of the imaginary part is simply due to sampling error and that hence in the limit of an infinite number of samples, the moments are real. 

In Figure \ref{fig3} we show a comparison of the mean and variance of species $X_1$ as predicted by the CLE-R1, CLE-R2 and CLE-C. The moments are normalized with the exact result obtained from the CME. The rate constants are $k_2 = k_3 = 1$. Note that the complex CLE (CLE-C) agrees within numerical error with the exact result from the CME, whilst significant deviations can be seen in the predictions of the real-valued CLEs (CLE-R1 and CLE-R2; see Section II C for definitions of these CLEs).

In Appendix F, we generalise the results of this section for any chemical system, derive the general properties of this CLE and in particular show that it does not break down for all times and that the moments of the complex variables are always real, i.e., it possesses a physically meaningful interpretation. The only properties used for this derivation are the analyticity and behaviour under complex continuation of the drift and diffusion terms in the CFPE - hence the generality of our approach. We also show in the same appendix that the moments of any analytic function in $\mathbb{R}^N$ and the autocorrelation functions and power spectra of the complex CLE are real-valued functions. We also therein discuss the method by which the complex CLE can be used to simulate first passage times.

\subsection{Applications}

Next we showcase the accuracy of the CLE-C for two systems of biochemical importance. In both cases we find the CLE-C's accuracy to be much higher than the accuracy of the conventional real-valued CLE as well as the accuracy of other popular methods in the literature. 

\subsubsection{The Michaelis-Menten reaction with substrate input}

We consider the Michaelis-Menten reaction with substrate input
\begin{align}\label{eq_e1}
  \varnothing & \xrightarrow{\quad c_4 \quad } S, \quad S+E \xrightleftharpoons[\quad c_2 \quad ]{c_1} C
  \xrightarrow{\quad c_3 \quad} E + X,
\end{align}
where $E$ is the free enzyme, $C$ is the enzyme-substrate complex, $S$ is the substrate and $X$ is the product. The number of enzyme molecules is fixed to one. The system has a steady state in the substrate concentration if $\alpha \equiv c_4 \Omega / c_3 < 1$, which simply means that the input rate must be slower than the maximum turnover rate. The CME for this reaction has been solved exactly in steady-state conditions (this has previously not been reported in the literature and hence we present a full derivation in Appendix G) leading to expressions for $P_0(n,\tau)$ and $P_1(n,\tau)$ - the probability of having $n$ substrate molecules at time $\tau$ given $0$ and $1$ free enzyme molecules, respectively. These are given by
\begin{align}
  P_0(n)
  & =
    \frac{C''}{k n!} \left(\frac{k_4}{k_3}\right)^{n+1}
    \frac{  \Gamma
     (n+k+1)}{\Gamma
     (k)} \,_1F_1 \left[ -n;-(k+n)
     ;k_3\right], \\
  P_1(n)
  & =
    \frac{C''}{n!} \left(\frac{k_4}{k_3}\right)^n
    \frac{  \Gamma
     (k+n )}{\Gamma
     (k)} \,_1F_1 \left[ -n;-(k+
     n-1); k_3 \right],
\end{align}
where $c=c_1/\Omega$, $k_2 = c_2/c, k_3=c_3/c, k_4= \Omega c_4/c$, $k = k_2 + k_4$, $k_{34} = k_3 / k_4$, $C''= e^{-k_4} (k_{34}- 1)^{k+1}/k_{34}^{k+1}$ and $\Omega$ is the compartment volume. The function $\,_1F_1$ is the confluent hypergeometric function. All moments of the fluctuations in the substrate and enzyme molecule numbers can thus be exactly computed without the need of stochastic simulation; this is convenient since it provides us with a means to rigorously check the accuracy of the CLE in standard form with reflective boundary conditions and of the complex CLE. 

The CLE in standard form (and the time rescaled by $c_1/\Omega$) is given by
\begin{align}\label{app_eq22}
  dx_1
  & = 
    (k_4 + k_2 ( 1-x_2) -  x_1 x_2) d\tau  
    + \sqrt{k_4} dW_1 + \sqrt{k_2 ( 1-x_2)} dW_2 -  \sqrt{ x_1 x_2} dW_3, \notag \\
  dx_2
  & = 
    ((k_2+k_3) ( 1-x_2) -  x_1 x_2) d \tau
    + \sqrt{k_2 ( 1-x_2)} dW_2 -  \sqrt{ x_1 x_2} dW_3
    + \sqrt{k_3 ( 1-x_2)} dW_4,
\end{align}
where $x_1$ is the number of substrate molecules and $x_2$ is the number of free enzyme molecules. Here we have used the conservation law between enzyme and complex molecules such that the number of complex molecules can be written as $1 - x_2$.

For simulations, we choose physiologically realistic values for the rate constants \cite{Fersht1999,Bareven2011}: $c_1=2 \times 10^6 (M s)^{-1}$, $c_2 = 1 s^{-1}$, and $c_3 = 
1 s^{-1}$. We choose the volume to be $\Omega = 10^6 M^{-1}$ which corresponds to a spherical submicron compartment of roughly $150$ nm diameter. The dimensionless parameter $\alpha \equiv c_4 \Omega / c_3$ is varied over the whole interval $[0,1]$ possible for steady-state through modification of the value of the input rate $c_4$. We simulate the system with the complex CLE (CLE-C) and a real version (CLE-R); in the latter we enforce the artificial boundaries $0 \leq x_1$ and $0 \leq x_2 \leq 1$ on the CLE in standard form Eq. (\ref{app_eq22}) (these boundaries ensure that the CLE-R does not break down; the upper boundary on $x_2$ reflects the fact that at any time the total amount of enzyme is at most one). 

Figure \ref{fig4} shows the mean and variance of both enzyme and substrate species obtained from the two CLEs normalized by the corresponding CME value (as determined from the exact solution - see Appendix G) as a function of $\alpha$. The results clearly show that the CLE-C's predictions for the first and second moments of the fluctuations of both species are of much higher accuracy than those of the CLE-R. For completeness sake, we have also compared the accuracy of these two CLEs with the modified CLE proposed by Dana and Raha (CLE-DR) \cite{Dana2011} and with two other popular methods in the literature: the Langevin equation obtained using the linear-noise approximation (LNA \cite{vanKampen,ElfEhrenberg}) and the two-moment approximation (2MA) which involves the closure of moment equations of the CME via the assumption of a negligible third cumulant \cite{GrimaJCP2012,Ferm2008,Ullah2009,Verghese2007}. The CLE-C gives significantly more accurate results than all three of these methods. We note that the 2MA gives significantly accurate results for the mean concentrations but not for the variances of fluctuations - this is in agreement with previous studies of the accuracy of moment closure methods \cite{GrimaJCP2012}. Given the four types of Langevin equations compared, the accuracy in ascending order is CLE-R and CLE-DR, LNA and CLE-C. The CLE-C is more accurate than the LNA because the latter is obtained from the CLE in the macroscopic limit \cite{GrimaThSt2011,Wallace2012}; however the LNA is more accurate than the CLE-R and CLE-DR because the latter suffer from artificially imposed boundaries or modified propensities to avoid its break down. 

Next, we consider the following first passage time problem. We want to compute the mean time it takes to produce a certain number of product molecules as a function of the initial substrate numbers. We explain in Appendix F how the complex CLE can be used to simulate first passage times. Figure \ref{fig4b} shows the mean first passage time $T$ for the catalytic reaction to produce $p_f = 10$ and $100$ product molecules. The values are normalized by those obtained from stochastic simulations using the SSA. The rate constants $c_1, c_2$ and $c_3$ and the volume $\Omega$ are chosen as before and in addition we set $c_4 = 10^5 Ms^{-1}$. We observe that the CLE-R gives much larger deviations from the SSA result than the CLE-C.

\subsubsection{A genetic negative feedback loop}

Next we consider a simple model of a genetic circuit with negative feedback
\begin{equation}\label{eq_g1}
  \begin{split}
  D_u & \xrightarrow{\quad r_u \quad } D_u + X, 
  \quad   X  \xrightarrow{\quad k_f \quad } \varnothing, 
  \quad D_u + X \xrightleftharpoons[\quad s_u \quad]{\quad k \quad } D_b.
\end{split}
\end{equation}
A gene in the unbound state $D_u$ expresses a protein $X$ which then acts to suppress its own expression through binding with the gene. In the bound state $D_b$ there is no production of protein. For simplicity the intermediate stage of mRNA production is ignored. The CME for this system has recently been solved exactly \cite{GrimaNewman2012}; hence as in the previous example, this system is ideal as a means to evaluate the closeness of the predictions of the various forms of the CLE to those of the  CME, while avoiding cumbersome stochastic simulations of the CME. 

The CLE in standard form (and with time rescaled by $k_f$) reads
\begin{align}
\label{app_eq42}
  dx_1
  & = 
    (-    x_1  +(\rho_u - \sigma_b  x_1) (1 - x_2) + \sigma_u x_2) d \tau \notag \\
  & \quad
    + \sqrt{\rho_u (1-x_2)}dW_1 -  \sqrt{x_1} dW_2 
    - \sqrt{\sigma_b x_1 (1-x_2)}dW_3  + \sqrt{\sigma_u x_2} dW_4, \notag \\
  dx_2
  & = 
    (\sigma_b  x_1 (1 - x_2)  - \sigma_u x_2) d \tau + \sqrt{\sigma_b x_1 (1-x_2)} dW_3 - \sqrt{\sigma_u x_2} dW_4,
\end{align}
where $x_1$ is the number of molecules of protein $X$, $x_2$ is the number of molecules of the bounded gene $D_b$, $\rho_u = r_u / k_f$, $\sigma_u = s_u / k_f$, $\sigma_b = k / (\Omega k_f)$ and $\Omega$ is the cellular volume. 

As before we implement the CLE in three different ways. The naive implementation enforcing reflective boundary conditions, i.e., $x_1 > 0$ and $0 < x_2 < 1$, such that the terms under the square roots in the standard form of the CLE Eq. (\ref{app_eq42}) remain positive (CLE-R), the complex version of the CLE (CLE-C) and the modified CLE of Dana and Raha \cite{Dana2011}. The simulations utilise the parameter set $r_u=10, k_f=1, s_u=0.5$. Figure \ref{fig5} shows the normalised mean number of molecules and the normalised variance of the protein and gene fluctuations as a function of the dimensionless parameter $\sigma_b=k/(\Omega k_f)$. This can be viewed as varying the bimolecular reaction constant $k$ for fixed volume $\Omega$ or equivalently as varying the volume of the system for fixed $k$. We observe a similar behaviour of the CLE-R and CLE-DR as for the enzyme system: their predictions are considerably more inaccurate than those of other methods (CLE-C, LNA and 2MA). The accuracy of the latter three methods is comparable though the CLE-C slightly outperforms the LNA and 2MA (see insets of Figure \ref{fig5}). 

\section{Conclusion}

Although the CLE is a popular and convenient analytical approximation to analyse stochastic chemical systems, nevertheless the issue of its boundary behaviour has been relatively neglected in the literature, despite well known problems arising in the low concentration limit. Here, we have shown that in general no boundary conditions can be enforced that maintain the accuracy of the CLE as an approximation to the CME while retaining real-valued state variables: hence the CLE (and CFPE) are only well defined as equations with complex number variables.

The main reason for this is that the CLE and CFPE confined to real space lead to a drift-diffusion process which is able to reach small enough values of the molecule numbers for which the diffusion matrix is not positive semi-definite, i.e., leads to a non well-defined CFPE (and CLE). In the macroscopic limit of large volume at constant concentrations, the probability that this happens is negligibly small because the drift-diffusion process is centered on very large molecule numbers and hence rarely approaches zero molecule numbers; this is why the CLE works well in this limit. In contrast the zero molecule number boundary is frequently visited whenever one or more chemical species have means of few molecule numbers which leads to ill-definition of the CLE in finite time. As we have shown, this problem is avoided by choosing the number variables in the CLE to be complex. This new CLE is always well-defined and hence does not suffer from break down. Its physical interpretation stems from the fact that it predicts real-valued mean concentrations and higher moments of intrinsic noise. We have also shown that the equivalence between the CLE and CME predictions for the mean concentrations and variance of fluctuations of unimolecular systems, a classical result using Ito calculus, is only obtained using the complex CLE. 

The complex nature of the CLE variables has been previously overlooked because the integrals leading to the moments of the CLE variables computed using Ito calculus do not need the precise specification of the domain of the CLE variables. This is the implicit reason why all attempts to generate a well-defined CLE drift-diffusion process in real space, using reflection boundary or drift / diffusion modification methods, lead to inaccurate predictions of the first two moments for unimolecular systems. 

We note that usually it is assumed that the CLE may lead to inaccurate results for systems with few molecule numbers \cite{RaoArkin2002} due to its implicit assumption of continuous molecule numbers,  rather than discrete. However as we have seen, simulation using the real-valued CLE requires the use of methods to artificially correct for its break down near the zero molecule number boundary, and hence the apparent inaccuracy of the CLE comes from the use of these methods as well as from its intrinsic assumption of continuous molecule numbers. When the CLE is considered in complex space, remarkably it is found to be accurate even for chemical systems with species in very low molecule numbers, such as the two bimolecular examples studied in Section III B, where the numbers of enzyme and gene were just one. This strongly suggests that the inherent inaccuracy of the CLE comes not so much from its assumption of continuous molecule numbers but rather from the methods used to correct for boundary effects if the domain of the CLE is assumed to be real. 

Our results also suggest that the complex CLE is particularly relevant to the simulation of biochemical systems since it is well known that such systems are typically characterised by many chemical species with few molecules per cell; for example in \emph{E. coli} the mean number of proteins per cell varies from 0.1 to about 1000, depending on the bacterial strain, with most strains exhibiting a mean protein number of 10 \cite{Taniguchi2010}. 

Our complex CLE is not the first use of stochastic differential equations in the complex plane to perform stochastic simulations of chemical processes. The only other such formalism is the Poisson representation (PR) developed by Gardiner and co-workers \cite{Gardiner2010,GardinerChaturvedi1977}. The stochastic differential equations in the PR are not an extension of the CLE, as in our present work. Rather they correspond to an exact Fokker-Planck equation in complex variables which is derived by an expansion of the probability distribution of the CME in Poisson distributions. The advantage of these stochastic differential equations over the complex CLE is that they are exact, i.e., all their moments are one and the same as those of the CME. Their main disadvantage is that their derivation requires the neglect of boundary terms in the process of integration which cannot be guaranteed and hence has to be checked on a case by case basis \cite{Gardiner2010}. The complex CLE does not suffer from such a problem and is generally applicable to all chemical systems; clearly its disadvantage compared to the PR stochastic differential equations is that it is an approximation of the CME. The latter is however not a significant issue in practice since as we have seen, the differences between the complex CLE and CME are typically small. 

The complex CLE involves the simulation of double the number of coupled stochastic differential equations as the conventional real-valued CLE, and hence it is typically found that more samples are needed to obtain accurate estimates of the moments.  Also in some cases, for example the enzyme and gene examples in Section III, we found that to guarantee numerical stability it was necessary to take a smaller time step size for the complex CLE than for the conventional CLE. Probably this restriction can be lifted or eased by use of more sophisticated stochastic differential equation simulation methods than the simple Euler-Maruyama one used in this article (see \cite{Higham} for a broad discussion of available methods). 

However we note that even using the Euler-Maruyama implementation, the complex CLE is computationally advantageous compared to the stochastic simulation algorithm whenever one is simulating systems characterised by many reactions per unit time and relatively few species. The complex CLE also achieves striking accuracy over a broad range of molecule numbers suggesting that it could be a novel useful tool in the chemical physicist's and computational biologist's arsenal. 

\section*{Acknowledgments}

We thank Philipp Thomas for interesting discussions. G.S. acknowledges support from the European Research Council under grant MLCS 306999.

\newpage

\appendix

\section{Break down analysis of the CLE for an isomerisation reaction}

The reaction is described by the scheme
\begin{align}
  X_1  \xrightleftharpoons[\quad c_2  \quad ]{c_1} X_2.
\end{align}
Rescale time $\tau= c_2 t$ and define the non-dimensional constant $k = c_1/c_2$. Due to the implicit conservation law in the total number of molecules of $X_1$ and $X_2$ this system is effectively unidimensional. The CLE in standard form for the number of molecules of species $X_1$ is given by
\begin{align}
  d x
  & = 
    ( - k x + (n_0-x)) dt - \sqrt{kx}~dW_1 + \sqrt{n_0-x}~dW_2,
\end{align}
where $n_0$ is the total number of molecules of $X_1$ and $X_2$. The terms in the square roots become negative for $x<0$ and $x>n_0$. Since one of the noise terms is always non-zero at the latter two values of $x$ it follows that the system can be driven to break down. An alternative form of the CLE is given by
\begin{align}
  d x
  & = 
    ( - (1+k) x + n_0) dt + \sqrt{(k-1)x + n_0}~dW.
  \end{align}
Let $b =(k-1)x + n_0$. For $k=1$, $b$ is always positive. For $k \ne 1$, $b = 0$ when $x= n_0/(1-k)$. At this point the drift causes $x$ to change according to the equation 
\begin{align}
  d x
  & = - \frac{2 k}{1-k} n_0 dt.
\end{align}
Thus for $k < 1$, $x$ decreases and consequently from the definition of $b$ one can see that $b$ increases above zero. Similarly for $k > 1$, $x$ increases and $b$ increases above zero as well. Hence for all values of $k$ we find that $b > 0$ for all times implying that the CLE in alternative form does not break down. Furthermore given that $b = 0$ when $x= n_0/(1-k)$ one can deduce that the value of $x$ in this CLE can become negative (for $k > 1$) or even exceed the value of $n_0$ (for $k < 1$); this is in contrast to the CME wherein the lower and upper bound of the number of molecules of species $X_1$ are 0 and $n_0$, respectively. 

\section{Break down analysis of the CLE for an open dimerisation reaction}

We consider the reaction system described by the scheme
\begin{align}
\varnothing \xrightarrow{\quad c_1 \quad} X_1, \quad 
  X_1 + X_1 \xrightarrow{\quad c_2 \quad} \varnothing.
\end{align}
Rescaling time as $\tau=t c_2/ \Omega$ and defining the non-dimensional constant $k= \Omega^2 c_1/c_2$, the CLE in standard form can be written as
\begin{align}
  dx 
  & = 
    (-2 x(x-1) + k) dt 
   -2 \sqrt{x(x-1) } dW_1 + \sqrt{k} dW_2.
\end{align}
The first noise term is zero for $x = 1$; the drift is positive in this case however the second noise term is non-zero and hence drives the system to $x < 1$, thus leading to the break down of this CLE.

An alternative form of the CLE is given by
\begin{align}
  dx 
  & = 
    (-2 x(x-1) + k) dt 
   + \sqrt{4 x^2 - 4x + k} dW.
\end{align}
Let $b = 4 x^2 - 4x + k$; this is positive for $k > 1$ and it becomes zero for $k \le 1$ for one of two $x$ values 
\begin{align}
 x_{\pm} 
 & = 
   \frac{1}{2}( 1 \pm \sqrt{1-k}).
\end{align}
Both values lie in the interval $[0,1]$. It is found that $b > 0$ for $x < x_-$ and $x > x_+$ while it is negative for $x_- < x < x_+$. Hence if $x(t = 0)$ is between $x_-$ and $x_+$ then the CLE will immediately break down.  The question is what happens if  $x(t = 0)$ is less than $x_-$ or larger than $x_+$. The drift $-2x(x-1)+k$ is positive in the interval $[0,1]$. Hence if $x(t = 0) < x_-$, the drift will lead to an increase in $x$ eventually causing this to take values in the interval $x_- < x < x_+$ for which $b < 0$; hence this case leads to a break down of the CLE in finite time. On the other hand if $x(t = 0) > x_+$, the drift will lead to an increase in $x$ in which case $b > 0$ and hence the CLE does not break down at any point in time.

\section{Break down analysis of the CLE for a closed dimerisation reaction}

The reaction is described by the scheme
\begin{align}
  X_1 + X_1 \xrightleftharpoons[\quad c_2 \quad ]{c_1} X_2.
\end{align}
We rescale time as $\tau=t c_2$ and define the non-dimensional constant $k =c_1/(\Omega c_2)$. The CLE in standard form is then given by
\begin{align}
  d x 
  & = 
    A(x) dt - 2\sqrt{k x(x-1)} dW_1 + 2\sqrt{\tfrac{1}{2}(n_0-x)} dW_2, 
\end{align}
where $A(x) = -2 k x(x-1) + (n_0 - x)$, $x$ is the number of $X_1$ molecules and $n_0$ is the maximum number of $X_1$ molecules (note the conservation law $x$ + 2$y$ = $n_0$ where $y$ is the number of $X_2$ molecules). One of the two noise terms is non-zero when the other one is zero and hence the noise causes $x$ to become less than 1 or greater than $n_0$ thus leading to the break down of the CLE.

An alternative form of the CLE is given by putting the two noise terms in one
\begin{align}
  d x 
  & = 
    A(x) dt + \sqrt{4 k x (x-1) + 2(n_0 -x)} dW. 
\end{align}
Let $D_1(x) = 4 k x (x-1) + 2(n_0 -x)$; then it follows that $D_1(x)$ becomes zero for the following two values of x
\begin{align}
 x_{\pm} 
 & = 
   \frac{2k+1 \pm \sqrt{4 k^2 + 4k(1-2n_0) + 1}}{4 k}.
\end{align}
These values are real provided $D_2(k)=4 k^2 + 4k(1-2n_0) + 1 > 0$. Now $D_2(k) = 0$ for the following two values of k 
\begin{align}
  k_{\pm}
  & = 
    (n_0-\tfrac{1}{2}) \pm \sqrt{(n_0-\tfrac{1}{2})^2 - \tfrac{1}{4}}.
\label{eqfork}    
\end{align}
The values $k_{\pm}$ are always real and positive because $n_0>1$. $D_2(k) > 0$ for $k \nin [k_-, k_+]$ and negative otherwise.  Hence we can state the following: (i) for $k \in [k_-, k_+]$, we have $D_2(k) < 0$ and thus there are no real values of $x$ for which $D_1(x) < 0$ which implies that the alternative CLE does not break down for all times and for all initial conditions.  (ii) for $k \nin [k_-, k_+]$, we have $D_2(k) > 0$ and thus $D_1(x) > 0$ for $x < x_-$ and for $x > x_+$ and $D_1(x) < 0$ for $x_-<x<x_+$. Hence if the initial condition is $x_-<x(0)<x_+$ then the CLE immediately breaks down. We next investigate what happens if $x(0) < x_-$ or $x(0)>x_+$ and $k \nin [k_-, k_+]$.

\subsection{Case $0 <k < k_-$}

We want to show that in this case $x_-> n_0$, $A(x_-)<0$ and $D_1(x_-)=0$ which means that the system does not break down. This is since only initial conditions with $x(0) \leq n_0$ are reasonable and the value of $x$ cannot increase above $x_-$ as the noise term is zero at this value and the drift then decreases $x$ below $x_-$ once this is reached.

We have 
\begin{align}
x_{-} (k)
& = 
  \frac{2k+1 - \sqrt{4 k^2 + 4k(1-2n_0) + 1}}{4 k}.
\end{align}
Since $n_0>1$ we have $16 k^2 n_0(n_0-1)>0$. Manipulating this we find
\begin{align}
 & \quad  16k^2 n_0^2 - 8 k n_0(2k+1)+4k^2+4k+1 >
4k^2+4k-8k n_0+1, \\
 \Leftrightarrow & \quad
   (2k+1 - 4 k n_0)^2 >
4 k^2 + 4k(1-2n_0) + 1.
\end{align}
Assume for now that $2k+1 - 4 k n_0 >0$. The right side of the equation is positive since its just the discriminant $D_2(k)$ defined above. We can thus take the square root of both sides to obtain
\begin{align}
 & \quad
   2k+1 - 4 k n_0 >
\sqrt{4 k^2 + 4k(1-2n_0) + 1}, \\
 \Leftrightarrow & \quad
 \frac{2k+1 - \sqrt{4 k^2 + 4k(1-2n_0) + 1}}{4 k}
> n_0.
\end{align}
The left side is equal to $x_-(k)$ and we have thus shown $x_-(k)>n_0$. We next show that $2k+1 - 4 k n_0 >0$ which was a necessary assumption for our proof. 

We can rewrite the term as $2k(1-2n_0)+1$ and see that it is monotonically decreasing in $k$ since $n_0 > 1$. Since we are restricted to the range $0 < k < k_-$, it is thus sufficient to show $x_-(k)>n_0$ for the maximal value of $k$, i.e.~for $k=k_-$. Since $n_0 > 1$ we have $n_0^2>n_0$. Algebraic manipulation gives 
\begin{align}
 & \quad
   5 n_0^2-n_0 > 4 n_0^2, \\
 \Leftrightarrow & \quad
   4 n_0^4-8n_0^3+5n_0^2-n_0 > 
4n_0^4-8n_0^3+4n_0^2, \\
 \Leftrightarrow & \quad
   (2 n_0 - 1)^2 (n_0^2-n_0)
>4 (n_0^2-n_0)^2.
\end{align}
Since both sides are positive we can take the square root of the last line and using $2 n_0 - 1>0$ and $n_0^2-n_0>0$ we obtain 
\begin{align}
 & \quad 
 (2 n_0-1) \sqrt{n_0^2-n_0} >
2 (n_0^2-n_0), \\
 \Leftrightarrow & \quad
 2 (2 n_0-1) \sqrt{n_0^2-n_0}  - 4 (n_0^2-n_0) > 0,\\
 \Leftrightarrow & \quad
   2 \left((n_0-\tfrac{1}{2}) - \sqrt{n_0^2-n_0} \right) (1-2n_0)+1 > 0, \\
 \Leftrightarrow & \quad
   2 k_- (1-2 n_0)+1>0, \\
 \Leftrightarrow & \quad
   2 k (1-2 n_0)+1>0,  
\end{align}
which verifies the assumption at the heart of the proof (see above) for $x_-(k)>n_0$. 
Finally, since $x_- > n_0 > 1$, we find $A(x_-)= -2 k x_-(x_- - 1) + (n_0 - x_-) < 0$.

We conclude that the system does not break down for $k < k_-$ independent of initial conditions.

\subsection{Case $k > k_+$}

We will show that for $k > k_+$ we have $0 < x_- <x_+ < 1$ and that the drift $A(x)$ is positive in the whole interval $[0,1]$. Recall that $D_1(x) > 0$ for $x < x_-$ and $x > x_+$ and negative otherwise. Hence if the initial condition is $x(0) > x_+$, the value of $x$ can decrease down to $x_+$ at which point $D_1(x_+) = 0$ and the drift is positive and hence $x$ increases above $x_+$ thus ensuring that $D_1$ is always positive and that no break down of the CLE occurs. However if the initial condition is $x(0) < x_-$ then the positive drift will cause $x$ to increase above $x_-$ in which case $D_1$ becomes negative and the CLE breaks down.  

The inequality $x_- <x_+$ is obvious from the definition of $x_{\pm}$. Starting from $0>- 8k n_0$ we obtain
\begin{align}
  4 k^2 + 4k+1 > 4 k^2 + 4k(1-2n_0) + 1,
\end{align}
where the right hand side is again $D_2(k)>0$. Since the left hand side is positive too, we can take the square root to obtain
\begin{align}
  & \quad 
    2k+1 > \sqrt{4 k^2 + 4k(1-2n_0) + 1}, \\
  \Leftrightarrow & \quad
    \frac{2k+1 - \sqrt{4 k^2 + 4k(1-2n_0) + 1}}{4 k} > 0, \\
  \Leftrightarrow & \quad
    x_- > 0.
\end{align}
Next, starting from $n_0>1$ we obtain
\begin{align}
  & \quad 
    8k(1-n_0) < 0, \\
  \Leftrightarrow & \quad
    4k^2+ 4k(1-2 n_0) + 1 < 4k^2 - 4k + 1,\\
  \Leftrightarrow & \quad
    4k^2+ 4k(1-2 n_0) + 1 < (2k-1)^2.
\end{align}
Since $n_0 > 0$, it follows from Eq. (\ref{eqfork}) that $k_+ > 1/2$. Thus the term in the parentheses on the right hand side of the above inequality is positive. The left side is again equal to $D_2(k)$ and thus positive. Taking the square root we find 
\begin{align}
  & \quad 
    \sqrt{4k^2+ 4k(1-2 n_0) + 1} < 2k-1, \\
  \Leftrightarrow & \quad
    \frac{2k+1 + \sqrt{4 k^2 + 4k(1-2n_0) + 1}}{4 k} < 1, \\
  \Leftrightarrow & \quad
    x_+ < 1.
\end{align}
Hence it follows from the above arguments that $0 < x_- < x_+ < 1$.

We next consider the drift for $x \in [0,1]$. First, consider the drift in the endpoints of this interval
\begin{align}
  A (0)
  & = 
    n_0 > 0, \\
  A(1)
  & = 
    n_0 - 1 >0. 
\end{align}
Since $A(x)$ is a parabola whose leading coefficient is negative, this means that $A(x) >0$ for all $x \in [0,1]$.

\section{Break down analysis of the CLE for a two enzyme catalysed reaction}
 
 Consider the system 
\begin{align}
  E_A + A  \xrightarrow{\quad c_1 \quad } E_A + B,
  \quad \varnothing \xrightleftharpoons[\quad c_4\quad ]{c_3} E_A, \notag \\
  \quad E_B + B  \xrightarrow{\quad c_2 \quad } E_B + A, 
  \quad \varnothing \xrightleftharpoons[\quad c_6 \quad ]{c_5} E_B.
  \label{enzymesys}
\end{align}
The enzymes are $E_A$ and $E_B$ and the substrates are $A$ and $B$. We here do not model the intermediate states of the enzyme and simply assume they are fast enough that we can ignore them. Define $N_0$, $x$, $x_A$, $x_B$ to be the total number of substrate molecules (those of $A$ and $B$), and the number of molecules of $A$, $E_A$ and $E_B$, respectively. Due to the implicit conservation law, the system is thus effectively a three variable one. 

Rescale time as $\tau = c_6 t$ and define $k_1= c_1/(\Omega c_6)  , k_2 = c_2/(\Omega c_6) , k_3 = \Omega c_3 / c_6, k_4=c_4/c_6, k_5=  \Omega c_5/c_6$. The CLE in standard form then reads

\begin{align}
  d x
  & = 
    (- k_1 x_A x + k_2 (N_0-x) x_B) d \tau - \sqrt{k_1 x_A x}~dW_1 + \sqrt{k_2 (N_0-x) x_B}~ dW_2, \\
  d x_A
  & = 
    (k_3-k_4 x_A)  d \tau + \sqrt{k_3}~dW_3 - \sqrt{k_4 x_A}~dW_4, \\
  d x_B
  & = 
    (k_5- x_B)  d \tau + \sqrt{k_5}~dW_5  - \sqrt{x_B}~dW_6 .
\end{align}

The CLE breaks down because when one of the noise terms is zero, the other noise term is not zero and hence the noise can drive the value of the variables such that the terms under the square roots are negative.  

An alternative form of the CLE is given by 
\begin{align}
  d x
  & = 
    (- k_1 x_A x + k_2 (N_0-x) x_B) d \tau + \sqrt{\lambda_1}~dW_1, \\
  d x_A
  & = 
    (k_3-k_4 x_A)  d \tau + \sqrt{\lambda_2}~dW_2, \\
  d x_B
  & = 
    (k_5- x_B)  d \tau + \sqrt{\lambda_3}~dW_3,
\end{align}
where 
\begin{align}
  \lambda_1
  & = 
    (k_1 x_A - k_2 x_B) x +  N_0 k_2 x_B , \\
  \lambda_2
  & = 
    k_3 + k_4 x_A, \\
  \lambda_3
  & = 
    k_5 + x_B.
\end{align}

Using Ito's formula \cite{Gardiner2010} and the above CLEs for $x_A$ and $x_B$, we can derive the CLEs for $\lambda_2$ and $\lambda_3$ leading to
\begin{align}
  d \lambda_2
  & = 
    k_4(2k_3 - \lambda_2)  d \tau + k_4\sqrt{\lambda_2}~dW_2, \\
  d \lambda_3
   & = 
    (2k_5- \lambda_3)  d \tau + \sqrt{\lambda_3}~dW_3.
\end{align}
Thus the CLEs in the variables $\lambda_2$ and $\lambda_3$ do not break down because when the noise terms equal zero (for $\lambda_2$ and $\lambda_3$ equal zero respectively), the drift terms become positive which leads to the eventual increase of the variables. This in fact could be deduced from our previous results as follows. The enzymes $E_A$ and $E_B$ are not influenced by the reactions involving $A$ and $B$. They simply undergo the simple birth and death process that has been investigated earlier (see Section II B) and whose alternative form CLE has been shown to not suffer from break down. 

Similarly we can deduce the CLE for variable $\lambda_1$ using the CLE for variable $x$ above. Under the constraint $k_1 x_A - k_2 x_B = \frac{k_1}{k_4}(\lambda_2-k_3) - k_2 ( \lambda_3 - k_5) \neq 0$, the new CLE reads
\begin{align}
  d \lambda_1
    & =  f(\lambda_1,\lambda_2,\lambda_3) d\tau 
    + \sqrt{\lambda_1} \left(\frac{k_1 (\lambda _2-k_3)}{k_4}+k_2 (k_5-\lambda _3) \right) dW_1
    + \sqrt{\lambda_2} k_1 \times \notag \\
    & \frac{k_4 (k_2 N_0 (k_5-\lambda _3)+\lambda _1)}{k_1 (\lambda _2-k_3)+k_2
   k_4 (k_5-\lambda _3)} dW_2 - \sqrt{\lambda_3} k_2 \frac{k_4 \lambda _1+k_1 N_0 (k_3-\lambda _2)}{k_1 (\lambda _2-k_3)+k_2
   k_4 (k_5-\lambda _3)} dW_3,
\end{align}
where $f$ is a complicated function of the variables $\lambda_1$, $\lambda_2$, and $\lambda_3$ and whose particular form is not important to our analysis. We find that as $\lambda_1$ becomes zero, the first noise term vanishes, however the other two noise terms are generally non-zero which implies that noise can drive $\lambda_1$ to negative values and hence the CLE breaks down. 

Thus, as for the two variable example in Section II C, the alternative form of the CLE does not circumvent the problems of the standard form of the CLE. Also similar to the results there, the break down is intimately related to the properties of the diffusion matrix of the CFPE. The diffusion matrix for the system (\ref{enzymesys}) is given by 
\begin{align}
  B
  & =
\left(
\begin{array}{ccc}
 \lambda_1 & 0 & 0 \\
 0 & \lambda_2 & 0 \\
 0 & 0 & \lambda_3 \\
\end{array}
\right).
\end{align}
Since this matrix is diagonal in the basis $x, x_A, x_B$, it follows that $\lambda_1$, $\lambda_2$ and $\lambda_3$ are its eigenvalues and that the matrix is positive semi-definite only if the eigenvalues are positive. Now the alternative form of the CLE breaks down at $\lambda_1 = 0$ which indeed corresponds to $B$ losing its positive semi-definite form and hence to an ill-defined CFPE. Hence the break down of all possible CLEs in real variable space for the enzyme system is guaranteed.  

\section{Positive semi-definiteness of the diffusion matrix associated with the CLE-C approach}

Let $C \in \mathbb{R}^{m \times n}$ be a real matrix,  $m,n \in \mathbb{N}$  and $B = C C^T \in \mathbb{R}^{m \times m}$. Let $\mathbf{v} \in \mathbb{R}^m, \mathbf{v} \neq 0,$ be an eigenvector of $B$ with eigenvalue $\lambda \in \mathbb{R}$ such that
\begin{align}
  B \mathbf{v}
  & = 
    \lambda \mathbf{v}.
\end{align}
For a vector $\mathbf{w} \in \mathbb{R}^p, p \in \mathbb{N}$, let $||\mathbf{w}||_p$ denote the Euclidean norm in $\mathbb{R}^p$, $||\mathbf{w}||_p = (\mathbf{w}^T \mathbf{w})^{1/2}$. Consider
\begin{align}
  \lambda ||\mathbf{v}||_m^2
  & = 
    \lambda \mathbf{v}^T \mathbf{v}
  =
    \mathbf{v}^T B \mathbf{v}
  =
    \mathbf{v}^T C C^T \mathbf{v}
  =
    (C^T \mathbf{v})^T (C^T \mathbf{v}) 
  =
    ||C^T \mathbf{v}||_n
  \geq
    0.
\end{align}
Since $\mathbf{v} \neq 0$, we have $||\mathbf{v}||_m^2 > 0$ and thus $\lambda \geq 0$. Since $B = C C^T$ is symmetric, it is diagonalizable. We have shown that all eigenvalues are non-negative and can thus conclude that $B$ is positive semi-definite. Conversely it follows that if the diffusion matrix $B$ is not positive semi-definite then the matrix $C$ cannot be real.

\section{General derivation of properties of the CLE-C}\label{extension_analytic}

Consider a Langevin equation of the form given by Eq. \eqref{eq6}. For the purpose of the following derivation we assume it to be in standard form, i.e. given by Eq. \eqref{eq8}. Now let the variables become complex such that the CLE reads
\begin{align}\label{eq_ext1}
  d \mathbf{z} 
  & = 
    \mathbf{A}(\mathbf{z}) dt + C(\mathbf{z}) \mathbf{dW}.
\end{align}
By writing $z_j=x_j +i y_j$ this equation can be split up into coupled Langevin equations for the real parts $x_j$ and imaginary parts $y_j$. By relabeling the variables as 
$(w_1, \ldots,  w_{2N})^T  =  (x_1, \ldots, x_N, y_1, \ldots ,y_N)^T$ we can write the equations in the form
\begin{align}\label{eq_ext2}
  d \mathbf{w}
  & = 
    \mathcal{A} dt + \mathcal{C} d \bf{ \mathcal{W}},
    \quad \mathcal{C} \mathcal{C}^T = \mathcal{B}.
\end{align}
This is the complex CLE (CLE-C). Here, we have defined

\begin{align}\label{eq_ext3}
  \mathcal{A}
  & = 
    (A_1^x, \ldots, A_N^x, A_1^y, \ldots, A_N^y)^T, 
    \quad 
  \mathcal{C} 
  = 
    \begin{pmatrix}
      C^x \\
      C^y 
    \end{pmatrix}, 
  \quad
  d\mathcal{W}
  =
     (dW_1, \ldots, dW_R) ^T. \\
  \mathcal{B}
  & = 
     \mathcal{C} \mathcal{C}^T
  =
    \begin{pmatrix}
      C^x (C^x)^T  &  C^x (C^y)^T \\
      C^y (C^x)^T  &  C^y (C^y)^T
    \end{pmatrix} 
   = 
    \begin{pmatrix}
      \mathcal{B}^{xx}  &  \mathcal{B}^{xy} \\
      \mathcal{B}^{yx}  &  \mathcal{B}^{yy}
    \end{pmatrix}.
    \label{dmatrixcvalued}
\end{align}

The superscripts $x$ and $y$ denote the real and imaginary part of a function $f(\mathbf{x},\mathbf{y})= f^x(\mathbf{x},\mathbf{y})+ i f^y(\mathbf{x},\mathbf{y})$. By construction the CLE-C does not contain square roots of negative expressions and is thus well-defined in $\mathbb{R}^{2N}$; hence it does not suffer from the break down problems of the common real-valued CLE. 

The new diffusion matrix $\mathcal{B}$ is symmetric; this follows from the fact that $\mathcal{B}^{xx}$ and $\mathcal{B}^{yy}$ are symmetric while $\mathcal{B}^{xy} = (\mathcal{B}^{yx})^T$. It is also the case that $\mathcal{B}$ is positive semi-definite since $\mathcal{C}$ is real in $\mathbb{R}^{2N}$ (see Appendix E). Note that the diffusion matrix of the CLE in real variables does not always possess this property over the real domain which indeed is intimately related to its break down as shown in Section II C in the main text. 

The corresponding FPE to the CLE-C reads 
\begin{align}\label{eq_ext4}
   \partial_t P(\mathbf{w},t)
   = &
     \Big[- \sum_{i=1}^{2N} \partial_i \mathcal{A}_i(\mathbf{w},t)
     + \tfrac{1}{2} \sum_{i,j=1}^{2N} \partial_i \partial_j
         \mathcal{B}_{ij}(\mathbf{w},t) \Big] P(\mathbf{w},t).
\end{align}
Next we show that this FPE is invariant under the operation $\mathbf{y} \rightarrow -\mathbf{y}$. First we rewrite the FPE Eq.~\eqref{eq_ext4} in the equivalent form
\begin{equation}\begin{split}\label{app_con_eq1}
  \partial_t P(\mathbf{x},\mathbf{y},t)
  & = 
    [- \sum_{i=1}^N(\partial_{x_i} A_i^x(\mathbf{x},\mathbf{y}) + \partial_{y_i}  A_i^y(\mathbf{x},\mathbf{y})) \\
  & \quad 
    + \tfrac{1}{2} \sum_{i,j=1}^N(\partial_{x_i}  \partial_{x_j} \mathcal{B}_{ij}^{xx}(\mathbf{x},\mathbf{y}) 
    + \partial_{y_i}  \partial_{y_j}  \mathcal{B}_{ij}^{yy}(\mathbf{x},\mathbf{y})
    + 2 \partial_{x_i}  \partial_{y_j}  \mathcal{B}_{ij}^{xy}(\mathbf{x},\mathbf{y}))] P(\mathbf{x},\mathbf{y},t),
\end{split}\end{equation}

where we have used $\partial_{x_i} \partial_{y_j} \mathcal{B}_{ij}^{xy}= \partial_{y_i} \partial_{x_j} \mathcal{B}_{ij}^{yx}$, which can easily be verified from the definition of $\mathcal{B}$ in \eqref{dmatrixcvalued}. 

Since $A_i$ is a polynomial with real coefficients, it fulfills $A_i(\bar{\mathbf{z}}) = \overline{A_i(\mathbf{z})}$ or in terms of $\mathbf{x}$ and $\mathbf{y}$ variables $A_i(\mathbf{x}, -\mathbf{y}) = \overline{A_i(\mathbf{x},\mathbf{y})}$. For its real and imaginary parts this implies 
\begin{align}
\label{Aprops}
 A_i^x(\mathbf{x},-\mathbf{y}) =  A_i^{x}(\mathbf{x},\mathbf{y}), \notag \\ A_i^y(\mathbf{x},-\mathbf{y}) =  -A_i^{y}(\mathbf{x},\mathbf{y}). 
\end{align} 
 
Using that $f_r$ are polynomials in the molecule numbers (see Section II A) and the symmetry properties of the complex square root, i.e.~$\sqrt{\bar{\mathbf{z}}} = \overline{\sqrt{\mathbf{z}}}$, we find $\sqrt{f_r(\bar{\mathbf{z}})}= \sqrt{\overline{f_r(\mathbf{z})}} = \overline{\sqrt{f_r(\mathbf{z})}}$. Given $C$ in the standard form, $C_{ir} = S_{ir} f_r^{1/2}$ this implies $C_{ir}(\bar{z}) = \overline{C_{ir}(\mathbf{z})}$. The real and imaginary parts thus obey $C_{ij}^x(\mathbf{x},-\mathbf{y}) = C_{ij}^x(\mathbf{x},\mathbf{y})$ and $C_{ij}^y(\mathbf{x},-\mathbf{y}) = -C_{ij}^y(\mathbf{x},\mathbf{y})$, respectively. Using these properties and the definition of $\mathcal{B}_{ij}$ given in Eq. (\ref{dmatrixcvalued}), it is straightforward to verify that 
\begin{align}
\label{Bprops}
\mathcal{B}_{ij}^{xx}(\mathbf{x},-\mathbf{y})=\mathcal{B}_{ij}^{xx}(\mathbf{x},\mathbf{y}), \notag \\ 
\mathcal{B}_{ij}^{yy}(\mathbf{x},-\mathbf{y})=\mathcal{B}_{ij}^{yy}(\mathbf{x},\mathbf{y}), \notag \\
\mathcal{B}_{ij}^{yx}(\mathbf{x},-\mathbf{y}) =  - \mathcal{B}_{ij}^{yx}(\mathbf{x},-\mathbf{y}).
 \end{align}
 
Using Eqs.~\eqref{Aprops} and \eqref{Bprops}, we find that the FPE in Eq.~\eqref{app_con_eq1} is invariant under the joint reflection of the imaginary variables, $\mathbf{y} \rightarrow -\mathbf{y}$ :
\begin{align}\label{app_con_eq2}
  \partial_t P(\mathbf{x},-\mathbf{y},t)
  & = 
    [- \sum_{i=1}^N(\partial_{x_i} A_i^x(\mathbf{x},-\mathbf{y}) + \partial_{-y_i}  A_i^y(\mathbf{x},-\mathbf{y})) \notag \\
  & \quad 
    + \tfrac{1}{2} \sum_{i,j=1}^N(\partial_{x_i}  \partial_{x_j} \mathcal{B}_{ij}^{xx}(\mathbf{x},-\mathbf{y}) 
    + \partial_{-y_i}  \partial_{-y_j}  \mathcal{B}_{ij}^{yy}(\mathbf{x},-\mathbf{y})
    + 2 \partial_{x_i}  \partial_{-y_j}  \mathcal{B}_{ij}^{xy}(\mathbf{x},-\mathbf{y}))] P(\mathbf{x},-\mathbf{y},t) \\
  & = 
    [- \sum_{i=1}^N(\partial_{x_i} A_i^x(\mathbf{x},\mathbf{y}) + \partial_{y_i}  A_i^y(\mathbf{x},\mathbf{y})) \notag \\
  & \quad 
    + \tfrac{1}{2} \sum_{i,j=1}^N (\partial_{x_i}  \partial_{x_j} \mathcal{B}_{ij}^{xx}(\mathbf{x},\mathbf{y}) 
    + \partial_{y_i}  \partial_{y_j}  \mathcal{B}_{ij}^{yy}(\mathbf{x},\mathbf{y})
    + 2 \partial_{x_i}  \partial_{y_j}  \mathcal{B}_{ij}^{xy}(\mathbf{x},\mathbf{y}))] P(\mathbf{x},-\mathbf{y},t).
\end{align}

Since the initial condition is always given by a symmetric probability distribution (the imaginary parts are necessarily zero since the initial specification is in terms of molecule numbers), it follows that the above invariance property implies that the probability distribution solution of the FPE Eq.~\eqref{app_con_eq1} remains symmetric for all times: $P(\mathbf{x}, \mathbf{y}, t) = P(\mathbf{x},-\mathbf{y},t)$. Finally we show that this implies real-valued moments of the complex variables in the FPE. 

Consider now a general moment 
$\langle z_1^{m_1} z_2^{m_2} \ldots z_N^{m_N} \rangle, m_1, \ldots m_N \in \mathbb{N}$,  
of the complex variables $z_i=x_i + i y_i$:
\begin{align}\label{app_con_eq3}
  \langle z_1^{m_1} \ldots z_N^{m_N} \rangle
  & =
    \int dz_1 \ldots dz_N~
    z_1^{m_1} \ldots z_N^{m_N} P(z,t) \notag \\
  & =
    \int dx_1 \ldots dx_N dy_1 \ldots dy_N~
    (x_1+ i y_1)^{m_1} \ldots (x_N + i y_N)^{m_N} P(x,y,t).
\end{align}

Each summand of the imaginary part of the product $(x_1+ i y_1)^{m_1} \ldots (x_N + i y_N)^{m_N}$ can be written in the form $x_1^{m_1-k_1} \ldots x_N^{m_N-k_N} y_1^{k_1} \ldots y_N^{k_N}$, with $k_i \in \mathbb{N}, k_i \leq m_i$ for $i=1, \ldots N$, and $\sum_{i=1}^N k_i$ is odd, i.e.~the exponents of the $y_i$ sum to an odd integer. The term $x_1^{m_1-k_1} \ldots x_N^{m_N-k_N} y_1^{k_1} \ldots y_N^{k_N}$ thus changes sign under $\mathbf{y} \rightarrow -\mathbf{y}$ and since the probability distribution is symmetric in $\mathbf{y}$ it then follows that the imaginary part of the integral in Eq.~\eqref{app_con_eq3} vanishes. This means that \emph{moments of the complex variables $z_i$ are real}.

Next, suppose we are interested in the moments of a general real-valued function $g(\bf{x})$. Suppose $g$ is analytic in $\mathbb{R}^N$ and that it can be globally represented as a power series.
This implies that it can be analytically continued to $\mathbb{C}^N$. Since $g(\bf{x})$ is real-valued the coefficients of a power series about a real point are real, too. This means that $g$ fulfills $g(\bar{\bf{z}}) = \overline{g(\bf{z})}$. As for the moments, since the probability is symmetric under $\mathbf{z} \to \bar{\mathbf{z}}$, the expectation of the imaginary part of $g(\mathbf{z})$ is zero, i.e., the expectation of $g(\mathbf{z})$ is real. Since powers of an analytic function have the same radius of convergence, the same also holds for all powers of $g$. This means that \emph{all moments of the function $g$ are real.}

Next, we consider the power spectrum of a stochastic process described by the complex CLE. The autocorrelation matrix for a homogeneous process can be computed by \cite{Gardiner2010}
\begin{align}\label{autocorr_def1}
  G(\tau)
  & = 
    \langle \mathbf{z}(\tau) \mathbf{z}^T(0) \rangle \\
  & = 
    \lim_{T \to \infty} \frac{1}{T} \int_0^T dt ~\mathbf{z}(t+ \tau) \mathbf{z}^T(t).
\end{align}
In terms of probability densities, it can be written as
\begin{align}
  G(\tau)
  & = 
    \int d \mathbf{z}_{\tau} d \mathbf{z}_0~\mathbf{z}_{\tau} \mathbf{z}^T_0 P( \mathbf{z}_{\tau}, \tau;  \mathbf{z}_0, 0),
\end{align}
where we defined $\mathbf{z}_t = \mathbf{z}(t)$. We have shown above that the solution of the FPE corresponding to the complex CLE is symmetric under the reflection of the imaginary variables, $\mathbf{y} \to -\mathbf{y}$ under appropriate initial conditions. It follows that transition probabilities and joint probability distributions have this property. Writing $\mathbf{z}_t = \mathbf{x}_t + i \mathbf{y}_t$ we have
\begin{align}
  \text{Im}[G_{ij}(\tau)]
  & = 
    \int d \mathbf{x}_{\tau} d \mathbf{y}_{\tau} d \mathbf{x}_0 d \mathbf{y}_0~((\mathbf{x}_{\tau})_i (\mathbf{y}_0)_j + (\mathbf{y}_{\tau})_i (\mathbf{x}_0)_j) P( \mathbf{x}_{\tau}, \mathbf{y}_{\tau}, \tau, \mathbf{x}_0, \mathbf{y}_0, 0).
\end{align}
The integrand is an odd function under the joint reflection $\mathbf{y}_{\tau} \to - \mathbf{y}_{\tau}, \mathbf{y}_0 \to -\mathbf{y}_0$, which means $\text{Im}[G_{ij}(\tau)]=0$, i.e., the correlation matrix $G$ is real. For a homogeneous process it further fulfills  $G(-\tau) = G(\tau)$ by construction. This means that the power spectrum, which is simply the Fourier transform of the autocorrelation matrix \cite{Gardiner2010}, is a real function given by
\begin{align}\label{power_spectrum}
  S(\omega)
  & = 
    \frac{1}{2 \pi} \int_{-\infty}^{\infty} d \tau e^{-i \omega \tau} G(\tau).
\end{align}
We thus have shown that the autocorrelation matrix can be obtained from the complex CLE using Eq.~\eqref{autocorr_def1}, leading to a \emph{well defined and real-valued power spectrum} via Eq.~\eqref{power_spectrum}.

Another physical quantity that is often of interest is the first passage time, i.e., the mean time the state vector $\mathbf{x}=(x_1, \ldots, x_N)$ takes to reach a particular value. For example, one may want to know the time it takes a certain number of protein molecules of some species to be produced. Say the molecule number of this species is labeled $x_i$; then the first passage time can be 
computed from the complex CLE by calculating the average time it takes for the real part of $x_i$ to achieve a certain value, i.e., we leave the imaginary parts unbounded.

\section{Exact solution of the CME describing catalysis by a single enzyme molecule}\label{app_enzyme}

Here we derive an exact solution to the CME for the enzyme reaction system described by scheme \eqref{eq_e1}. To the best of our knowledge this has not been previously reported; a previous exact derivation led only to explicit expressions for the mean substrate concentration \cite{Stefanini2005}. 

Define $c=c_1/\Omega$, rescale time as $\tau = c t$ and define $k_2 = c_2/c, k_3=c_3/c, k_4= \Omega c_4/c$. Let $P_0(n,\tau)$ and $P_1(n,\tau)$ be the probability of having $n$ substrate molecules at time $\tau$ given $0$ and $1$ free enzyme molecules, respectively. The coupled CME's describing the time evolution of these two probabilities are then given by
\begin{align}\label{app_eq21}
  \partial_{\tau} P_0(n, \tau)
  & = 
      k_4 P_0(n-1, \tau) 
    +   (n+1) P_1(n+1, \tau) 
    -   k_4  P_0(n, \tau)
    -    k_2 P_0(n, \tau) 
    -    k_3 P_0(n, \tau), \\
\label{app_eq21b}
  \partial_{\tau} P_1(n, \tau)
  & = 
      k_4 P_1(n-1, \tau) 
      +   k_2  P_0(n-1, \tau) 
      +   k_3  P_0(n, \tau)
      -   k_4 P_1(n, \tau)
      -   n P_1(n, \tau). 
\end{align}
Define the generating functions as
\begin{align}\label{app_eq23}
  G_0(s)
  & = 
    \sum_{n} s^{n} P_0(n), \\
  G_1(s)
  & = 
    \sum_{n} s^{n} P_1(n).
\end{align}
Multiplying \eqref{app_eq21} and \eqref{app_eq21b} in steady state ($\partial_{\tau} P_0 = \partial_{\tau} P_1 = 0$)  with $s^n$ and summing over $n$ leads to
\begin{align}\label{app_eq24}
  0
  & = 
    (k_4(s-1)-k_2-k_3) G_0(s) +  \partial_s G_1(s), \\
  0
  & = 
    (k_2 s + k_3)  G_0(s) + (k_4(s-1) -  s \partial_s) G_1(s). 
\end{align}
Solving the second equation for $G_0$ and inserting into the first gives
\begin{align}\label{app_eq25}
  0
  & = 
     \frac{(k_4(s-1)-k_2-k_3)k_4(1-s)}{k_2 s + k_3}  G_1(s) + \left( \frac{(k_4(s-1)-k_2-k_3)  s}{k_2 s + k_3}+ 1 \right) \partial_s G_1(s),
\end{align}
which leads to the solution
\begin{align}\label{app_eq26}
  G_0(s)
  & = 
    C \frac{e^{k_4 s}}{(k_3 - k_4 s)^{k_2+k_4+1}}
  =
    C' \frac{e^{k_4 s}}{(k_{34} - s)^{k+1}}, \\
  G_1(s)
  & = 
    C \frac{e^{ k_4 s}}{(k_3- k_4 s)^{k_2+k_4}}
  =
    C' \frac{e^{k_4 s}}{(k_{34} - s)^{k}},
\end{align}
where $k_{34}= k_3 / k_4$, $k=k_2 + k_4$, and $C'$ is a normalization constant. The latter can be obtained from the normalisation condition
\begin{align}\label{app_eq27}
   \sum_n (P_0(n)+P_1(n))
  = 
    G_0(1) + G_1(1) 
  =  1,
\end{align}
which leads to
\begin{align}\label{app_eq28}
  C'
  & = 
    \frac{e^{-k_4}}{k_{34}} (k_{34}- 1)^{k+1}.
\end{align}
Hence the generating function solution is given by
\begin{align}
  G_0(s)
  & = 
     \frac{e^{k_4 (s-1)}}{k_{34}} \left(\frac{k_{34} - 1}{k_{34} - s} \right)^{k+1} , \\
  G_1(s)
  & = 
     \frac{e^{k_4 (s-1)}}{k_{34}} \frac{(k_{34} - 1)^{k+1}}{(k_{34} - s)^{k}}.
\end{align}
One can show by induction that
\begin{align}
\label{Gfsol1}
\partial_s^n G_0(s) = 
  & = 
    \frac{C'}{k} \frac{\sum_{i=0}^n \binom{n}{i} [k]^{n-i+1}(k_4 (k_{34}-s))^i}{(k_{34}-s)^{k+n+1} } e^{k_4 s} \notag \\
  & = 
    \frac{(k_{34}- 1)^{k+1}}{k_{34} k} \frac{e^{k_4 (s-1)}}{(k_{34}-s)^{k+n+1} }
     \sum_{i=0}^n \binom{n}{i} [k]^{n-i+1}(k_4 (k_{34}-s))^i , \\
\label{Gfsol2}
 \partial_s^n G_1(s) = 
  & = 
    C' \frac{\sum_{i=0}^n \binom{n}{i} [k]^{n-i}(k_4 (k_{34}-s))^i}{(k_{34}-s)^{k+n} }e^{k_4 s} \notag \\
  & = 
    \frac{(k_{34}- 1)^{k+1}}{k_{34}} \frac{e^{k_4 (s-1)}}{(k_{34}-s)^{k+n} }
     \sum_{i=0}^n \binom{n}{i} [k]^{n-i}(k_4 (k_{34}-s))^i ,
\end{align}
where we have used the definition for the rising factorial
\begin{align}
  [k]^i
  & = 
    k \cdot (k+1) \ldots (k+i-1), \\
  [k]^0
  & =  
    1.   
\end{align}
The probability distribution functions can now be obtained using their definition in terms of the generating functions
\begin{align}
\label{pexact0}
  P_0(n)
  & = 
    \frac{1}{n!}\partial_s^n G_0(s)|_{s=0}, \\
\label{pexact1}
  P_1(n)
  & = 
    \frac{1}{n!}\partial_s^n G_1(s)|_{s=0}.
\end{align}
Substituting Eqs. (\ref{Gfsol1}-\ref{Gfsol2}) in Eqs. (\ref{pexact0}-\ref{pexact1}), leads to
\begin{align}
  P_0(n)
  & = 
    \frac{C''}{k} \frac{k_4^{n+1}}{n!} \sum_{i=0}^n \binom{n}{i} [k]^{n-i+1}k_3^{i-n-1},  \\
  P_1(n)
  & = 
    C'' \frac{k_4^n}{n!} \sum_{i=0}^n \binom{n}{i} [k]^{n-i}k_3^{i-n},
\end{align}
where $C''= e^{-k_4} (k_{34}- 1)^{k+1}/k_{34}^{k+1}$. These can be compactly represented in terms of the 
confluent hypergeometric function $_1F_1$ which leads to our final solution of the CME for the enzyme reaction system
\begin{align}\label{app_eq31}
  P_0(n)
  & =
    \frac{C''}{k n!} \left(\frac{k_4}{k_3}\right)^{n+1}
    \frac{  \Gamma
     (n+k+1)}{\Gamma
     (k)} \,_1F_1 \left[ -n;-(k+n)
     ;k_3\right], \\
  P_1(n)
  & =
    \frac{C''}{n!} \left(\frac{k_4}{k_3}\right)^n
    \frac{  \Gamma
     (k+n )}{\Gamma
     (k)} \,_1F_1 \left[ -n;-(k+
     n-1); k_3 \right].
\end{align}
Analytic expressions for moments of arbitrary order can be directly computed by taking appropriate derivatives of the generating functions in Eqs. (G14) and (G15).
The average number of substrate molecules $\langle n \rangle$, the average number of enzyme molecules $\langle n_E \rangle$ and the variance in fluctuations about these averages ($\Sigma$ for the substrate and $\Sigma_E$ for the enzyme) are thus given by
\begin{align}\label{app_eq35a}
    \langle n \rangle
  & = 
    \frac{k_4 k_3 (k_3 + k_2) + k_4^2   }
    {  k_3(k_3-k_4)}, \\
  \langle n_E \rangle 
  & =
    1 - \frac{k_4}{k_3}, \\
  \Sigma
  & = 
\frac{k_4 (k_3^2 (k_4^2+k_3
   (-k_4+k_2+k_3))+k_4  
   (k_3
   (k_4+k_3)-k_4^2))}{ 
   k_3^2 (k_4-k_3)^2}, \\
  \Sigma_{E}
  & = 
    \frac{k_4(k_3-k_4)}{k_3^2}.
\end{align}

\newpage

\newpage

\newpage

\newpage

\begin{figure}[t]
  \includegraphics[scale=.4]{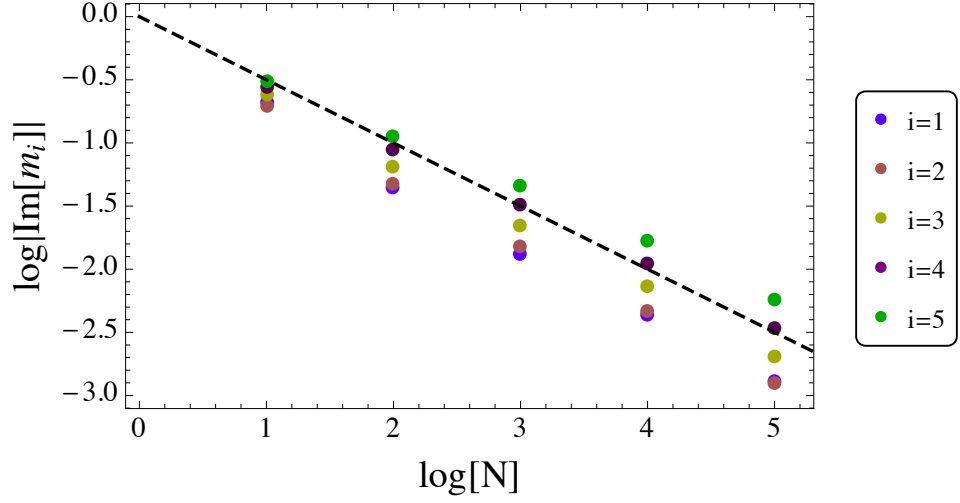} 
  \caption{Absolute value of the imaginary part of the first five moments of species $X_1$ in reaction \eqref{eq9} as a function of the number of samples $\mathrm{N}$ in CLE-C simulations. The $i^{th}$ moment is given by $m_i$. Each value is normalized by the absolute value of the corresponding moment. The points of each moment can be approximately fitted by a straight line (black dashed curve) with a slope of $-1/2$ . This implies that the normalized imaginary parts decay as $\sim 1/\sqrt{\mathrm{N}}$ and thus converge to zero in the limit of an infinite number of samples. The simulation parameters are $k_1 = 0.3, k_2 = k_3 = 1$, $\delta \tau = 10^{-3}, \Delta \tau = 1$.
  }
  \label{fig2}
\end{figure}

\newpage

\begin{figure}[t]
  \includegraphics[scale=0.5]{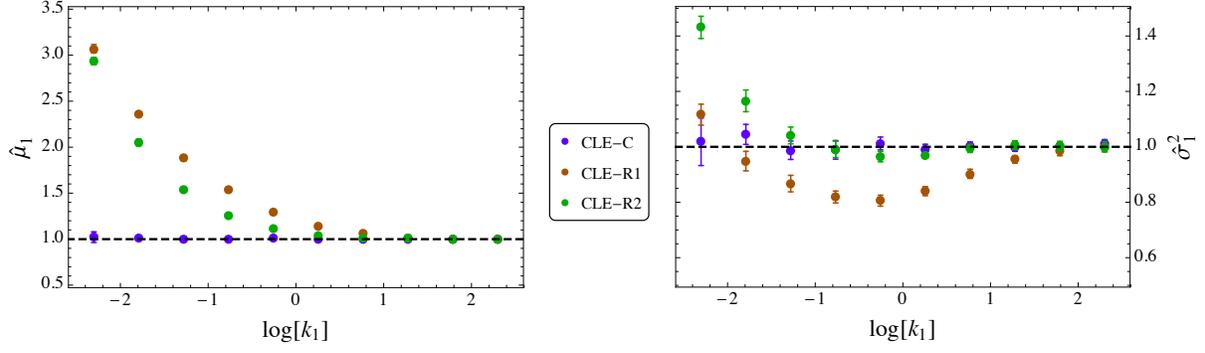}
  \caption{Testing the accuracy of the complex CLE (CLE-C). This is the same plot as in Figure \ref{fig1}, but with the result of the CLE-C included. The CLE-C gives the correct mean and variance of species $X_1$, i.e, agrees with the CME, within sampling error. The CLE-C's superior accuracy over that of the real-valued CLEs stems from the fact that the CLE-C does not suffer from break down and that hence it does not need the imposition of artificial boundaries (as necessary for the real-valued CLE-R1 and CLE-R2). Simulation details for all three implementations are $\delta \tau = 10^{-3}, \Delta \tau = 1, N = 10^4$.}
  \label{fig3}
\end{figure}

\newpage

\begin{figure}[t!]
\centering
  \includegraphics[scale=0.5]{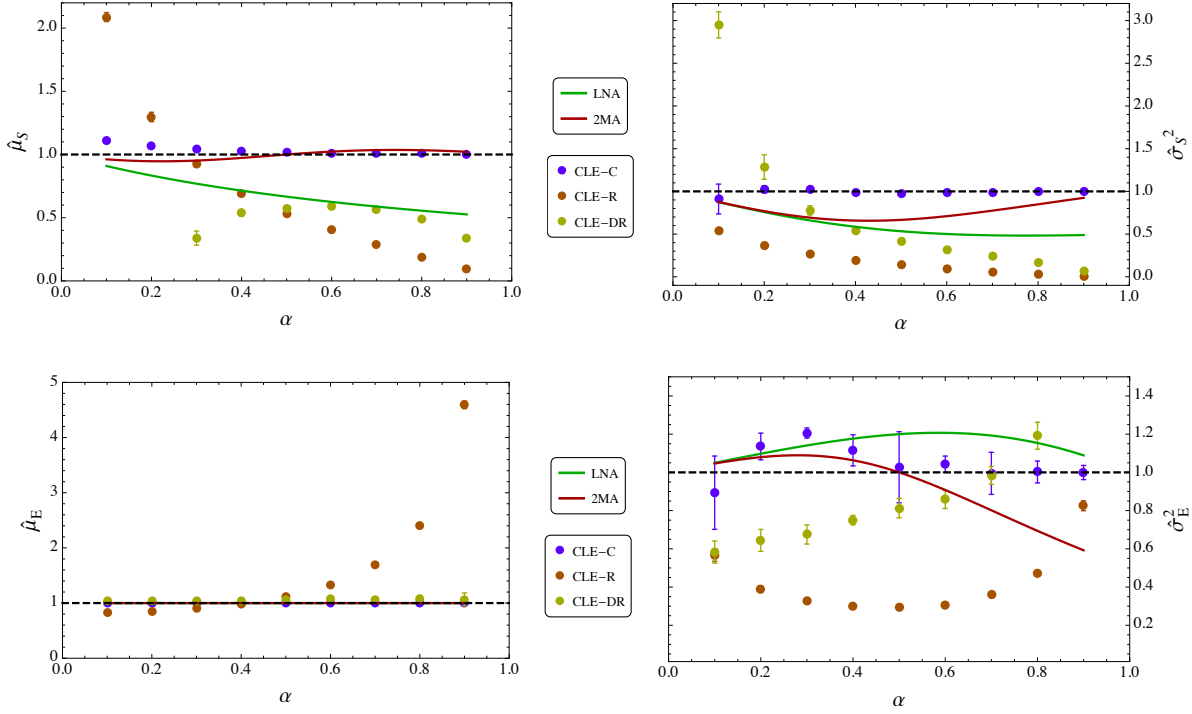}
\\
  \caption{Normalised mean number of molecules of substrate $\hat{\mu}_S$ and enzyme $\hat{\mu}_E$, and their corresponding variances $\hat{\sigma}_S^2$ and $\hat{\sigma}_E^2$, as a function of the non-dimensional parameter $\alpha$ (a measure of saturation), for the enzyme reaction system \eqref{eq_e1}. The values are normalized by the exact values obtained from the CME which are derived in Appendix G. We find that the CLE-R (CLE in standard form with artificial reflective boundaries to avoid break down) and CLE-DR (a modified CLE proposed in  \cite{Dana2011}) give generally worse results than the CLE-C. The latter is also significantly more accurate than both the conventional LNA and the 2MA approximations. The simulation parameters are as follows. For the CLE-C: $\delta \tau = 10^{-4}, N=10^5$. $\Delta \tau$ scales like $\alpha^4$ from $5-45$ for $\alpha =0.1 - 0.9$; for the CLE-R and CLE-DR: $\delta \tau =10^{-4}$, $\Delta \tau =10$ and $N=10^4$.}
  \label{fig4}
\end{figure}

\newpage

\begin{figure}[t!]
\centering
  \includegraphics[scale=0.5]{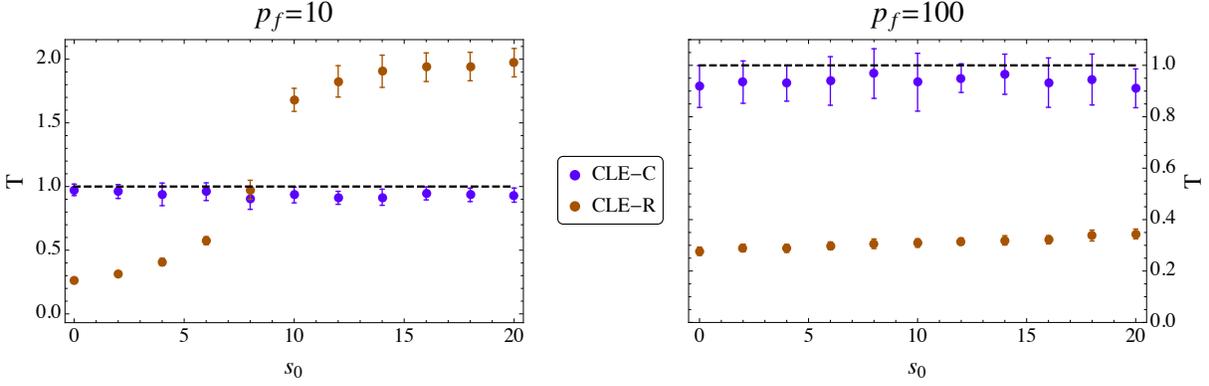}
  \caption{Normalised mean first passage time $T$ for a number $p_f$ of product molecules to be produced, as a function of the initial substrate concentration $s_0$, for the enzyme reaction system \eqref{eq_e1}. The values are normalized by the exact values corresponding to the CME obtained by stochastic simulations using the SSA. We find that the CLE-R (CLE in standard form with artificial reflective boundaries to avoid break down) gives generally worse results than the CLE-C. For the simulation time step we used $\delta \tau = 10^{-3}$ for the CLE-C and CLE-R. The number of samples drawn were $10^3$ and $10^2$ for $p_f = 10$ and $100$ respectively.}
  \label{fig4b}
\end{figure}

\begin{figure}[t!]
\centering
  \includegraphics[scale=0.5]{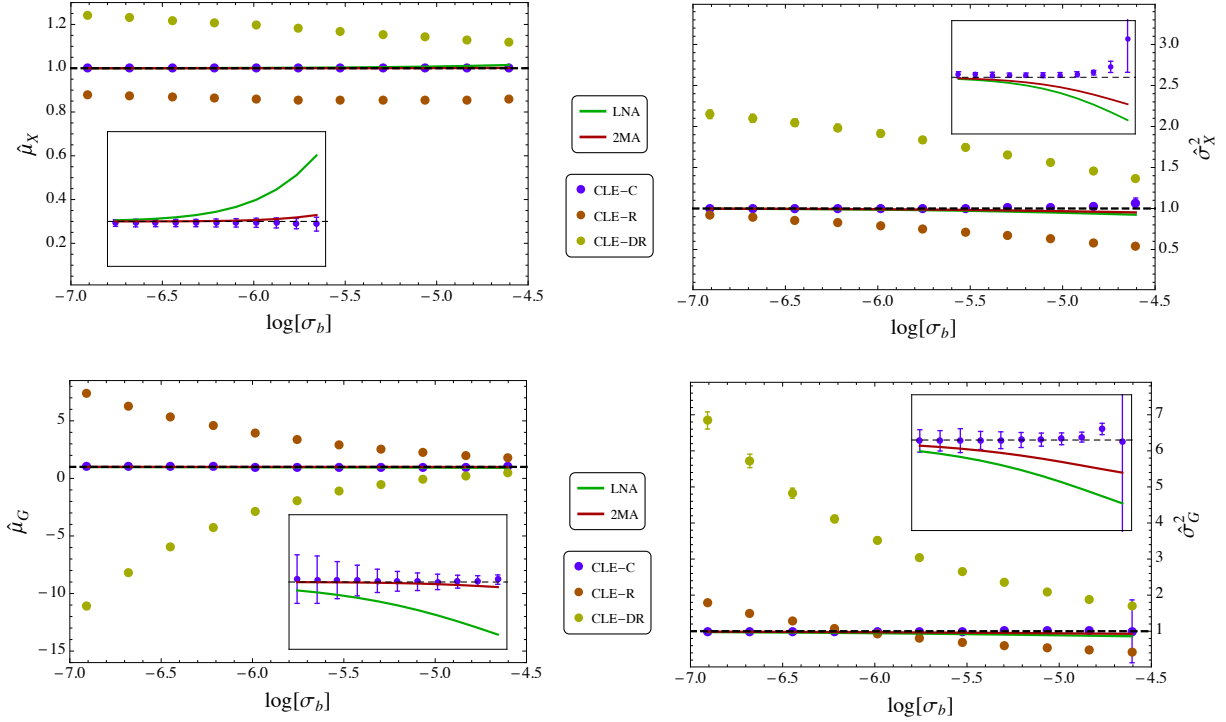} 
  \caption{Normalised mean number of molecules of protein $\hat{\mu}_X$ and bound gene $\hat{\mu}_G$, and their corresponding variances $\hat{\sigma}_X^2$ and $\hat{\sigma}_G^2$, as a function of the non-dimensional parameter $\sigma_b$ (a measure of binding affinity of the protein to the gene), for the genetic negative feedback loop \eqref{eq_g1}. The values are normalized by the exact values obtained from the CME \cite{GrimaNewman2012}. We find that the CLE-R (CLE in standard form with artificial reflective boundaries to avoid break down) and CLE-DR (a modified CLE proposed in \cite{Dana2011}) give generally worse results than the CLE-C. The accuracy of the latter and of the conventional LNA and the 2MA approximations are comparable. The simulation parameters are as follows. For the CLE-R and CLE-DR: $\delta \tau = 10^{-4}, \Delta \tau =10, N=10^4$. For the CLE-C: $\delta \tau = 10^{-4}, \Delta \tau =10, N=10^5$.}
  \label{fig5}
\end{figure}

\end{document}